\documentclass[letterpaper]{article} % DO NOT CHANGE THIS
\usepackage{aaai25}  % DO NOT CHANGE THIS
\usepackage{times}  % DO NOT CHANGE THIS
\usepackage{helvet}  % DO NOT CHANGE THIS
\usepackage{courier}  % DO NOT CHANGE THIS
\usepackage[hyphens]{url}  % DO NOT CHANGE THIS
\usepackage{graphicx} % DO NOT CHANGE THIS
\usepackage{natbib}  % DO NOT CHANGE THIS AND DO NOT ADD ANY OPTIONS TO IT
\usepackage{caption} % DO NOT CHANGE THIS AND DO NOT ADD ANY OPTIONS TO IT
\usepackage{algorithm}
\usepackage{algorithmic}
\usepackage{tabularx}
\usepackage{booktabs}
\usepackage{caption}
\usepackage{subcaption}
\usepackage{newfloat}
\usepackage{listings}
\DeclareCaptionStyle{ruled}{labelfont=normalfont,labelsep=colon,strut=off} % DO NOT CHANGE THIS
\floatstyle{ruled}
\newfloat{listing}{tb}{lst}{}
\floatname{listing}{Listing}
\title{A Human-Centered Approach to Identifying Promises, Risks, \& Challenges of Text-to-Image Generative AI in Radiology}
\author {
    Katelyn Morrison\textsuperscript{\rm 1},
    Arpit Mathur\textsuperscript{\rm 1},
    Aidan Bradshaw\textsuperscript{\rm 1},
    Tom	Wartmann\textsuperscript{\rm 2},
    Steven Lundi\textsuperscript{\rm 3},
    Afrooz Zandifar\textsuperscript{\rm 4},
    Weichang Dai\textsuperscript{\rm 5},
    Kayhan Batmanghelich\textsuperscript{\rm 5},
    Motahhare Eslami\textsuperscript{\rm 1}\equalcontrib,
    Adam Perer\textsuperscript{\rm 1}\equalcontrib
}
\affiliations {
    \textsuperscript{\rm 1}Carnegie Mellon University\\
    \textsuperscript{\rm 2}ETH Zurich\\
    \textsuperscript{\rm 3}UCLA Health\\ 
    \textsuperscript{\rm 4}UPMC\\ 
    \textsuperscript{\rm 5}Boston University\\
    kcmorris@andrew.cmu.edu, arpitmam@andrew.cmu.edu, abradsha@andrew.cmu.edu, wd2119@bu.edu, batman@bu.edu, meslami@andrew.cmu.edu, adamperer@cmu.edu
}
\usepackage{bibentry}
\begin{document}

\maketitle

\begin{abstract}
    As text-to-image generative models rapidly improve, AI
    researchers are making significant advances in developing domain-specific 
    models capable of generating complex medical imagery from text prompts. Despite 
    this, these technical advancements have overlooked whether and how medical professionals would benefit from and 
    use text-to-image generative AI (GenAI) in practice.
    By developing domain-specific GenAI without involving stakeholders, we risk the potential of building models that are either not useful or even more harmful than helpful. In this paper, we adopt a human-centered approach to responsible model development by involving stakeholders in evaluating and reflecting on the promises, risks, and challenges of a novel text-to-CT Scan GenAI model. Through exploratory model prompting activities, we uncover the perspectives of medical students, radiology trainees, and radiologists on the role that text-to-CT Scan GenAI can play across medical education, training, and practice. This human-centered approach additionally enabled us to surface technical challenges and domain-specific risks of generating synthetic medical images.
    We conclude by reflecting on the implications of medical text-to-image GenAI. 
\end{abstract}


\section{Introduction}

The rapid integration of Generative Artificial Intelligence (GenAI) into various workplace and educational settings, including healthcare, has simultaneously 
sparked excitement and concern. 
GenAI holds great promise for augmenting workflows, improving productivity, and enhancing training across a wide range of industries~\cite{chan2023comprehensive,ooi2023potential}. 
However, the rapid development of domain-specific GenAI without understanding the needs of and challenges faced by domain stakeholders raises the risk of GenAI being misused.
Within healthcare, there is a growing amount of literature on new architectures for text-to-image (T2I) GenAI capable of generating complex medical imagery from text prompts~\cite{mai2024application,hamamci2023generatect,xu2024medsyn,chen2024medical}. However, much of the research has focused on technical advancements, such as novel architectures, instead of being rooted in the needs and challenges of medical stakeholders.
% However, key stakeholders have had limited involvement, leaving us with an incomplete understanding of the promises and challenges of these models in practice. %, such as anatomical diagrams~\cite{noel2024evaluating,zhu2024can}, CT scans~\cite{hamamci2023generatect,xu2024medsyn}, chest X-rays~\cite{chen2024medical}, and disease-relevant phenomena~\cite{koljonen2023could,ooi2023potential,waisberg2023text}.

Traditionally, synthetic image generation has been used to address pivotal technical medical imaging tasks~\cite{sorin2020creating}, such as image reconstruction or correction~\cite{levac2024accelerated}.
% Now, with foundation T2I models, such as DALL-E~\cite{ramesh2021zero} and MidJourney~\cite{midjourney2023}, becoming more easily accessible to end-users, 
% technical health-specialized researchers bring light to the limitations of using foundation models to generate medical images from text prompts%for use in medical education and decision-making processes
% ~\cite{zakka2020mammoganesis,kumar2024using,mai2024application,noel2024evaluating,waikel2023generative,waisberg2023text,zhu2024can,hamamci2023generatect,muhr2024evaluating,koljonen2023could,yang2020xraygan}.
% , furthing limiting our understanding of the promises and perils of T2I GenAI in applied contexts. 
Now, with foundation T2I models becoming widely accessible, the allure of simulating rare conditions, creating patient vignettes, or augmenting training datasets is attracting significant interest~\cite{thesen2024ai,perez2025radedit,waikel2023generative,waisberg2023text}. However, much of this enthusiasm is driven by technological evaluations~\cite{noel2024evaluating,bhardwaj2023notion,lin2024evaluating,ku2023viescore} rather than human-centered ones, leading to a \textit{disconnect} from key stakeholders' real needs and challenges. As recently emphasized by human-centered AI researchers, evaluations need to extend beyond algorithmic metrics by understanding how the AI can address stakeholders' real needs and what challenges arise by doing so~\cite{sivaraman2025over}. Although recent research captures medical stakeholders' perspectives on GenAI's potential applications when used in practice~\cite{yildirim2024multimodal}, few works capture medical stakeholders' perspectives on the technical challenges and domain-specific risks of generating synthetic medical imagery. 

Several studies have identified and evaluated the challenges and risks of domain-agnostic T2I models~\cite{bird2023typology,naik2023social,bianchi2023easily}. However, it is unclear how these challenges and risks translate to domain-specific T2I models. To address these gaps in literature, we follow up a quantitative evaluation of a novel medical T2I model~\cite{xu2024medsyn} with a human-centered qualitative evaluation. As an interdisciplinary team with expertise in human-computer interaction, machine learning, and radiology, we conducted a human-centered evaluation of the promises, technical challenges, and domain-specific risks of a text-guided model that generates Computed Tomography (CT) images of the lung (text-to-CT Scan GenAI)~\cite{xu2024medsyn}. CT scans provide radiologists with cross-sectional views of the internal structures of the human body to aid in diagnosis~\cite{mayo_ct_scan}. 

Our human-centered approach in this work consists of two phases (formative discussions and model exploration) to facilitate discussions about: (\textbf{RQ1}) How do key stakeholders imagine using text-to-CT Scan GenAI for medical education, training, and practice? and (\textbf{RQ2}) What technical and domain challenges emerge from stakeholders' needs and workflows for medical text-to-CT Scan GenAI? To address these questions, we engaged with a total of eight radiologists, four trainees, and two medical students. We will refer to this group as key stakeholders throughout the paper.

Our formative discussions included six radiologists and one senior trainee, and they took place at the end of the quantitative evaluation of the model in the study conducted by~\cite{xu2024medsyn}. These discussions focused on capturing stakeholders' ideas on how the model might be applied in practice (\textbf{RQ1}).
% The discussions lasted for as long as the remaining session time allowed and primarily focused on capturing stakeholders' ideas on how the model might be applied in practice (\textbf{RQ1}). 
This informed the type of tasks we gave participants during the second phase (model exploration), broadened our network for recruiting additional participants, and informed the selection of participant demographics. 
For the second phase, we recruited two medical students, three radiology trainees, and four radiologists (including two from the formative discussions) to prompt and evaluate the model in a semi-structured interview format. To facilitate easy prompting and output review, participants interacted with the model through a plugin that we developed for an existing open-source medical imaging interface (created by~\citet{ohif_viewer}). By interacting with the model, participants imagined scenarios of how it can address real needs and challenges faced throughout radiology workflows (\textbf{RQ1}). The exploratory prompting activity also shed light on the technical challenges and potential domain-specific risks that researchers should consider when developing medical T2I models (\textbf{RQ2}).

Our work makes three contributions to the AIES and broader Responsible AI (RAI) community. First, we present a text-to-CT Scan GenAI plugin for a popular open-source medical imaging viewer (created by \citet{ohif_viewer}), which researchers can extend to explore additional approaches for human-centered evaluations of the ethical and safety challenges of domain-specific T2I models. 
Second, we are the first work to leverage a human-centered approach to explore a medical T2I model with medical stakeholders. As a result of this, our paper expands upon~\citet{yildirim2024multimodal}'s human-centered GenAI work for applications of multimodal AI in radiology by mapping out applications of domain-specific T2I across medical education, radiology training, and practice.
Third, we extend existing RAI work and build on~\citet{munuera2023generative}'s position on the implications of using medical T2I GenAI by presenting technical challenges and domain-specific risks that emerged from participants interacting with the model. These challenges and risks range from topics such as confirmation bias, misrepresentation of image findings, and output image resolution. We discuss the implications of developing medical T2I and suggest future research directions to consider exploring.

%%% ------------- RELATED WORKS ---------- %%%%%
\section{Background and Related Works}

We distinguish the novelty of our contributions by highlighting the latest research on human-centered approaches to evaluating text-to-image GenAI. Additionally, we contextualize our work by synthesizing the current literature that has discussed the promises and challenges of medical text-to-image GenAI in education and practice.

\subsection{Human-Centered Approaches to T2I Evaluation}
% this subsection of related works is about human-centered model evaluation of T2I

The development and deployment of GenAI systems has attracted significant attention, especially as GenAI becomes widely accessible across domains, such as healthcare, education, and creative industries. As they become accessible across domains, researchers have set out to understand and mitigate harm from and misuse of T2I models~\cite{gu2024survey,naik2023social,bianchi2023easily,bird2023typology}. 
% Human-centered design principles have long emphasized aligning system functionalities with user needs to ensure interactive designs, emphasizing iterative feedback and transparency, regardless of the domain~\cite{amershi2019guidelines}. 
As part of this effort, numerous works have designed benchmarks and metrics to help quantify the harm and safety of T2I models~\cite{lee2023holistic,libenchmarking,hao2024harm}. \citet{parrish2023adversarial} and \citet{deng2025weaudit} create frameworks and tools to engage the public in auditing the safety of T2I models.~\citet{collins2024beyond} design a feedback mechanism with thumbs up/down as a way to operationalize human-centered evaluations.~\citet{lin2024evaluating} and~\citet{ku2023viescore} leverage image-to-text GenAI to quantitatively evaluate model generations through prompt and text output alignment. 
% \citet{evirgen2024text} conducted a user study with crowdworkers using a T2I model to identify different types of transparency mechanisms that support users' model understanding. 
\citet{bird2023typology} conducted a literature review to synthesize the different types of risks that emerge for researchers to consider when developing and evaluating domain-agnostic T2I models. Human-centered design principles, agnostic to the type of GenAI application, have since been developed, emphasizing the importance of co-creativity, transparency, and generative variability~\cite{weisz2024design,chen2023next}.

Several studies have recently evaluated the capability of domain-agnostic foundation models to generate medical images~\cite{bhardwaj2023notion,moin2024assessment,temsah2024art}.~\citet{noel2024evaluating} and~\citet{kumar2024using} similarly probe different text-to-image GenAI models to evaluate their capability in generating different medical images. One study ran an educational intervention study with students to quantitatively evaluate the impact of generative AI images in education about pediatric genetics~\cite{waikel2023generative}.~\citet{thesen2024ai} recruited a diverse group of stakeholders to evaluate generated images for education across several scales, including the images' potential to be used in medical presentations.
The above-mentioned studies primarily focus on quantitative evaluations of domain-agnostic T2I models with proxy metrics, such as anatomical realism or the presence of diseases, instead of assessing whether T2I capabilities meaningfully address stakeholders' needs and challenges.
Our work builds on existing studies by uncovering the promises, risks, and challenges of T2I capabilities in the context of medical education, training, and practice through prompting activities with relevant stakeholders.

\subsection{Promises and Challenges of Medical T2I GenAI}

\textbf{In Medical Education.} With text-to-image models increasing in competency and becoming widely accessible, researchers are exploring the possibility of using these models to generate medical images for education and training, according to a recent survey~\cite{mai2024application}. In fact, the use of foundation models to generate medical images is becoming so popular among the medical community that~\citet{safadi2024advancing} created a guide for medical educators on using generative AI in presentations. 
% As such, we identify medical education (the time that students are in medical school or residency) as a viable application context for this work.

Numerous previous works from medical literature have explored foundation models' ability to generate images sufficient to use for medical education~\cite{waikel2023generative,kumar2024using,koljonen2023could,zhu2024can,noel2024evaluating}.  
 For example,~\citet{zhu2024can} probe the potential of OpenAI's DALL-E~\cite{ramesh2021zero} to generate ECGs, including ECGS that are abnormal for teaching purposes. Although they do not present a study on probing or interviewing stakeholders, the authors use their professional experiences to provide insight into DALL-E's potential as a teaching tool. 
 % Ultimately, they deemed DALL-E incompetent at generating medically relevant ECG illustrations.
  \citet{currie2024generative} raise numerous concerns about using T2I models in nuclear medicine for education, citing the need to identify policies and regulations that address misuse and abuse, misrepresentations and hallucinations, and patient privacy concerns. Our paper adds to these findings by capturing perspectives on T2I in medical education from a diverse group of medical stakeholders.

\textbf{In Clinical Practice.}
Clinical decision-making informed by medical imaging is a nuanced process that relies on expertise, peer collaboration, iterative review, and the use of reference resources to augment diagnostic accuracy and patient care. 
Despite advancements in AI, existing tools and interfaces tend to overlook these needs and focus on higher-order problems such as pathology detection~\cite{Sloan2024} or automated report writing~\cite{yousefirizi2022aibaseddetectionclassificationpredictionprognosis}. 
However, recent work has argued for the need to explore AI's true value to clinicians throughout their clinical workflows~\cite{sivaraman2025over}. A recent study by~\citet{waisberg2023text} looks at text-to-image GenAI's potential as a collaborative tool for clinicians in helping them understand neuro-visual diseases' impact on patients' ability to see by generating an image based on a description of the patient's symptoms. However, no formal quantitative or human-centered evaluations were conducted. \citet{munuera2023generative} suggests how T2I can be used by physicians to simulate treatment effects to navigate treatment plans or by radiologists to enhance noisy images. The authors simultaneously cite general concerns and potential implications of misuse and data bias. 

 Although previous works have identified risks with T2I models in medical education, training, and practice, they are primarily based on domain-agnostic T2I models, such as DALL-E or MidJourney. Furthermore, they primarily consist of showing already generated images to participants instead of allowing key stakeholders to explore and prompt the model themselves. Our work builds on the existing literature by having medical students, radiology residents, and radiologists interact with a domain-specific T2I to capture richer applications and challenges of medical T2I.

%%% ------------- METHODS ---------- %%%%%
\section{Study Design}
\label{probe}

 We present the main takeaways from our formative discussions and then describe the protocol for our semi-structured interview. We also provide technical details on the text-to-CT Scan plugin that participants used during the interview. 
 % Our study aims to address two research questions: (\textit{\textbf{RQ1}}) \textit{How do key stakeholders imagine using a medical T2I GenAI tool?} (\textit{\textbf{RQ2}}) \textit{What AI safety and domain-specific challenges emerge as key stakeholders prompt and explore generations from a medical T2I GenAI?}

 \begin{figure*}
    \centering
    \includegraphics[trim={0cm 26cm 0cm 0cm},clip,width=0.7\textwidth]{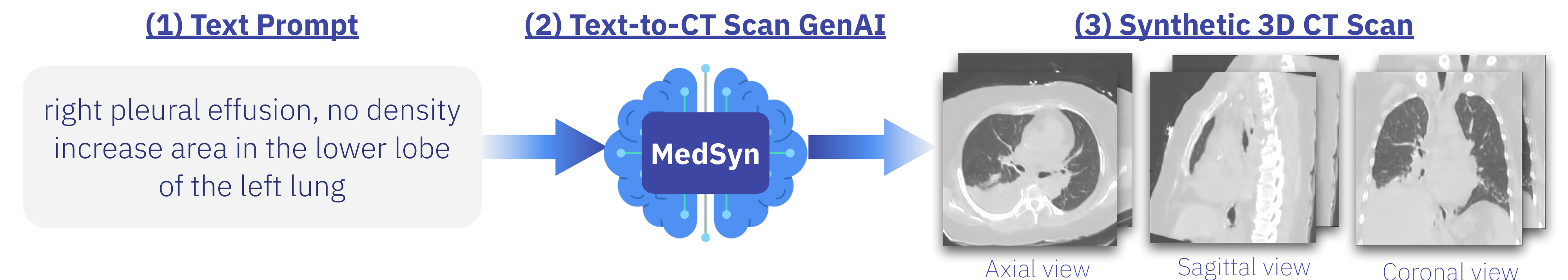}
    \caption{\textbf{Input/Output of the Text-to-CT Scan GenAI Model~\cite{xu2024medsyn}.} The text-to-CT Scan GenAI model that we use, called MedSyn~\cite{xu2024medsyn}, takes in a text prompt from the user to generate a 3D CT Scan. \textit{Brain icon by Icons8.}}
    \label{fig:mode-overview}
\end{figure*}

\begin{figure*}[h!]
    \centering
    \includegraphics[trim={0cm 6cm 3cm 0cm},clip,width=0.85\linewidth]{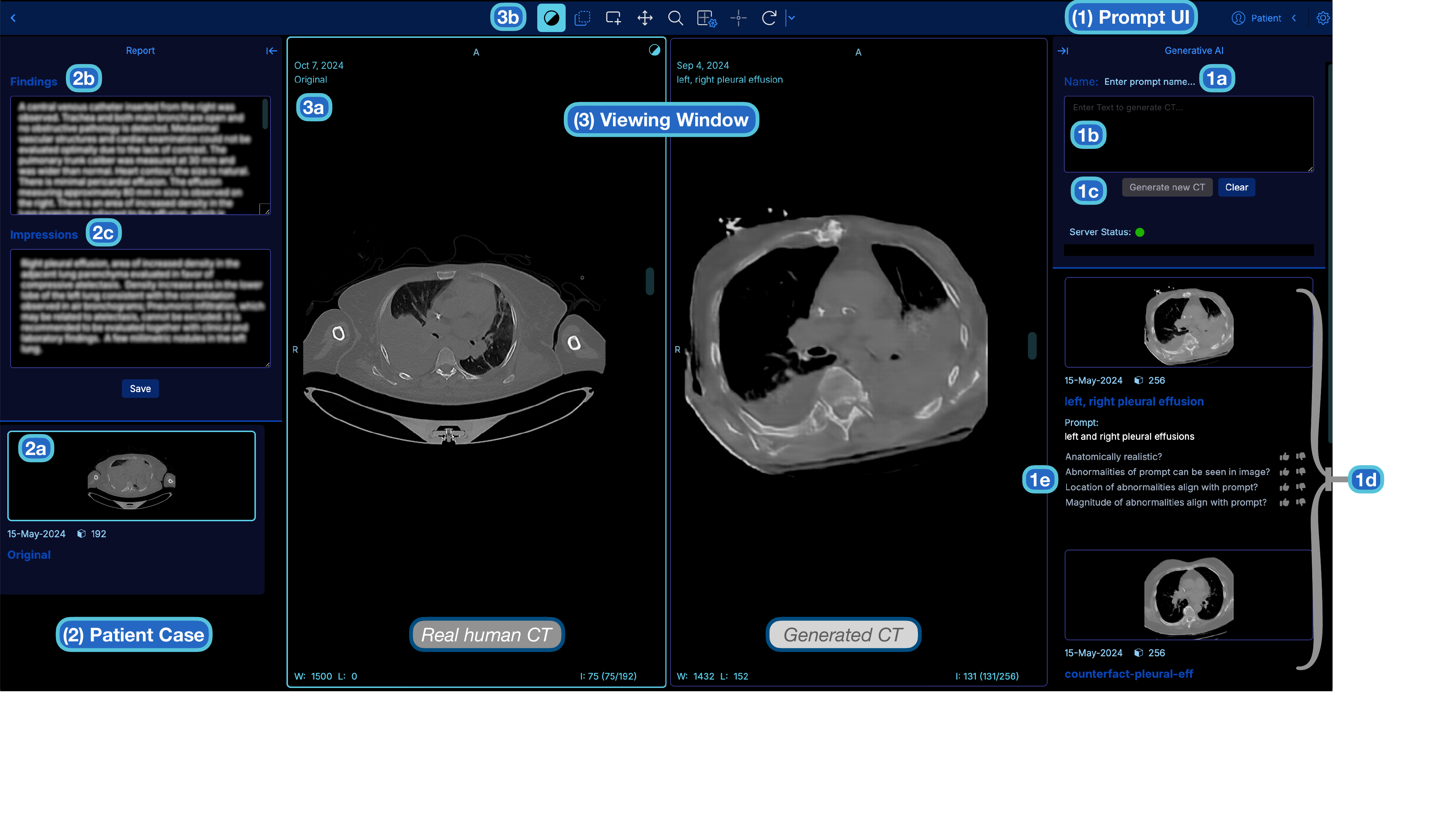}
    \caption{\textbf{Open-sourced medical imaging interface by~\citet{ohif_viewer} with our GenAI plugin powered by MedSyn~\cite{xu2024medsyn}.} \textbf{(2) Patient Case.} The first column shows a real case consistent with severe right pleural effusion (sampled from CT-RATE dataset~\cite{hamamci2023generatect,hamamci2024foundation,hamamci2024ct2rep}. \textbf{(3) CT Scan Viewing Window.} The second and third columns show a CT scan comparison viewing window, where a human's CT scan (from~\cite{hamamci2023generatect,hamamci2024foundation,hamamci2024ct2rep}) is on the left, and a generated CT scan from MedSyn~\cite{xu2024medsyn} is on the right. \textbf{(1) Prompt UI.} The fourth column is our text-to-CT Scan GenAI plugin, allowing users to generate a CT scan. }
    \label{fig:design-probe}
\end{figure*}

\subsection{Formative Discussion Takeaways} 
% Summarize the formative study in one paragraph with the main points -- describe the formative studies - the methods, and participants, don't show those.
We interviewed six radiologists and one senior radiology trainee (F1-F7) for an average of $6.8$ minutes after they quantitatively evaluated outputs from the MedSyn model in~\citet{xu2024medsyn}'s study\footnote{Provided in Appendix Section Formative Study}. Ultimately, the formative discussions provided us with preliminary insights into radiologists' challenges and their perceptions of the value of text-to-CT Scan GenAI in education, training, and practice. However, participants struggled to detail specific scenarios since many had never used T2I GenAI before. We also uncovered participants' challenges with current online resources, such as visualizing the exact pathologies they search for, the importance of visual comparative analyses, and lack of exposure to T2I GenAI. These challenges inspire us to integrate MedSyn into the medical imaging interface by~\citet{ohif_viewer}.

\subsection{Text-to-CT Scan GenAI Plugin}

 Our formative discussions with radiologists revealed to us that they frequently search for image findings and pathologies on domain-specific search engines. However, they may not always find exactly what they are searching for. Therefore, we developed a GenAI plugin powered by the text-to-CT Scan GenAI model, MedSyn~\cite{xu2024medsyn}, and integrated it into a popular open-source medical image viewer~\cite{ohif_viewer}\footnote{Plugin Demo available at https://genai-radiology.web.app/.}. The MedSyn model enables users to generate CT scans based on exact and specific pathologies, extending capabilities beyond current search tools. As depicted in Figure\textbf{~\ref{fig:mode-overview}} below, MedSyn takes a user's text prompt and generates a 3D CT Scan consisting of the axial, sagittal, and coronal views of the lungs. Generating the complete CT scan takes between three and four minutes when using an A100 GPU. Our plugin manifests in the interface (shown in Figure\textbf{~\ref{fig:design-probe}} below) as a text box where users can (\textbf{1a}) name their prompt, (\textbf{1b}) write their prompt, and (\textbf{1c}) generate a CT scan by calling the model API through the click of a button. Users can view a list of previous CT scans that they generated (\textbf{1d}) and have the opportunity to provide feedback on the generation (\textbf{1e}). 
 % We host the design probe online.

\subsubsection{Front-end Interactions.} 
% We use the Open Health Imaging Foundation \cite{ohif_viewer} framework as the foundation for our design probe, as seen in Figure\textbf{~\ref{fig:design-probe}}. 
When using this tool, a radiologist can view a patient's case history (\textbf{2a}), noting their findings (\textbf{2b}) and impressions (\textbf{2c}). This functionality is incorporated to align with how radiologists currently read CT scans in clinical practice. Viewing the patient's CT in the main viewer (\textbf{3a}) gives them access to numerous functionalities (\textbf{3b}) that they are used to having, such as adjusting the contrast to preset window levels (\textit{e.g.}, soft tissue, bone, and lung)\footnote{Presets were determined by an author, who is a radiologist.}, adding annotations, comparing images side-by-side, and using a localizer tool. 
Many of these functionalities either expand upon what the existing radiology search engines offer or leverage findings from our formative interviews. For example, the tool allows users to view and compare a generated CT scan with a real CT scan, as well as annotate them. 

\subsection{Semi-structured Interviews}
\label{interviews}
We conducted 45-minute semi-structured interviews with relevant medical stakeholders to elicit their perspectives on the potential applications and implications of text-to-CT Scan GenAI. These interviews were designed to be exploratory in nature; the goal was not to formally audit or analyze prompting behaviors. The interview sessions consist of four parts, inspired by~\citet{cai2019hello}'s approach with a clinical decision-making design probe, which included a \textit{pre-probe}, \textit{probe}, and \textit{post-probe} phase. We describe each part in the following subsections.

\subsubsection{Introduction.} The interviews started with an introduction and background discussion, where participants shared their experiences and challenges throughout their medical education, training, and practice. This portion was designed to understand the tools, methods, and resources they use to learn about different pathologies and image findings in CT scans. 

\subsubsection{Pre-probe.}
\label{preprobe} Participants were asked about their experiences with text-to-image GenAI. For participants who have never used text-to-image GenAI before, we showed them a short conversation made in ChatGPT where we already requested the generation of a medical image from DALL-E\footnote{Provided in Appendix Section Pre-Probe Materials.}. All participants were asked before interacting with the model if they could think of any use cases for text-to-image GenAI being valuable to them throughout their education, training, or practice to prime them for the rest of the interview session. Following this, we showed all participants an animated illustration of how the text-to-CT Scan GenAI model works, and participants were acclimated to the interface through a guided walkthrough video.
% before interacting with the interface themselves.

\subsubsection{Probe.} During the probe phase, participants generated two to four %$2$--$4$ 
CT scans, depending on session duration and the depth of discussions. They were guided to use the model to generate CT scans based on different goals informed by the formative interviews, such as validating their impression of a real patient's CT scan loaded in the viewer from CT-RATE~\cite{hamamci2023generatect,hamamci2024foundation,hamamci2024ct2rep}, learning about a pathology through differential diagnosis, and generating something of their choice. Throughout this process, participants provided feedback about the prompts they chose, the quality of the generation, and insights on the model's potential applications in radiology education, training, and practice. 

\subsubsection{Post-probe.} Finally, after interacting with the model, participants were asked to reflect on the risks of generating synthetic CT scans in real-world applications and raise any concerns about the generated output.

\subsection{Participant Recruitment}% \arpit{
Participants were recruited through snowball sampling methods from our clinical collaborators' networks. For each session, we provided the participants with an overview of the study's purpose and methods and obtained their consent for participation. All participants consented to audio and screen recording for data analysis and were provided a form of monetary compensation for their time. We interviewed nine stakeholders (R1--R9) across three different institutions: two medical students, three radiology trainees,
% \footnote{Residency programs in the United States come after someone has graduated from medical school, allowing them to gain further practice and experience in a specialization of their choice. In this work, we refer to residents as trainees.}, 
and four attending radiologists (two from the formative study and two new ones); experience distribution is provided in Table~\ref{tab:phase2_demographics}). Sessions were held on Zoom or Teams, based on preference, and lasted for an average of $46.34$ minutes.

\begin{table}[t]
    \centering
    \begin{tabular}{cccc}
        PID & Expertise Level & Radiology Experience (years) \\
        \hline
        R1  & Expert   & 5--10 \\
        R2  & Trainee  & 1     \\
        R3  & Expert   & $>20$   \\
        R4  & Expert   & 5--10 \\
        R5  & Expert   & $>20$   \\
        R6  & Trainee  & 1     \\
        R7  & Student  & 0*    \\
        R8  & Student  & 0**   \\
        R9  & Trainee  & 1     \\
        \hline
    \end{tabular}
    \caption{Participant Demographics for interview sessions. *Fourth-year med student. **Third-year med student.}
    \label{tab:phase2_demographics}
\end{table}
% } 

\subsection{Analysis}

We used an affinity diagramming approach to analyze the qualitative data. First, the transcripts from the interview sessions were verified for accuracy, and corrections were made by listening to the recordings as necessary. Quotes were pulled out from the transcripts, and affinity mapping was conducted by multiple authors to group similar quotes across interviews based on common themes. All quotes were grouped into primary themes and sub-themes that aligned with our research questions. Multiple authors grouped the quotes individually, and then the final groups were determined through discussion.

%%% ------------- RESULTS ---------- %%%%%

\section{Findings}
%\section{Design Probe Exploration and Evaluation}

% [introduction to our opportunities and challenges findings]

% \arpit{
We first report on the identified promises of T2I in medical education, training, and practice %by synthesizing our conversations and the participants' prompts 
(\textbf{RQ1}). We then follow up on these promises with the risks and challenges of T2I  GenAI that stakeholders identified (\textbf{RQ2}). Table\textbf{~\ref{tab:participant_applications}} and Table\textbf{~\ref{tab:participant_challenges}} show high-level representations of the findings for both research questions. A complete list of the prompts created by each participant, along with their comments about the prompt and the generated output, is provided in the supplemental materials.
%that emerged from this analysis.
% In this section, we present findings organized along three high-level themes, namely (1) Design Opportunities in Radiology Education and Training, (2) Design Opportunities in Radiology Practice, and (3) Challenges for Designing Medical Text-to-3D Image GenAI. These findings present important considerations and opportunities for future research on the adoption of text-to-image generation in medicine and radiology specifically.
% }

\begin{table*}
\centering
\begin{tabular}{|c|l|c|}
\hline
\textbf{Phase} & \textbf{Promises of Text-to-CT Scan GenAI} & Participants \\ 
\hline
Education \& Training & Supplement Lectures \& Practice Materials  &  F2, R1, R3, R7, R8\\ 
  & Learning through Variations, Differentials, \& Rare Cases  & F6, R1, R2, R6, R9 \\ 
  & Empower Trainees when On-Call or in Emergency Situations & F3, R2, R7, R8\\
  \hline
Practice  & Improving Report Impression of Medical Images & F5, R1, R4--R6 \\ 
 & Planning \& Communicating with Others & R2, R6, R7 \\
& Visual Memory Support & F6, R4, R5  \\ 
\hline
\end{tabular}
\caption{(\textit{RQ1}) Participants mapped to which applications of text-to-CT Scan GenAI they imagined across the different phases.}
\label{tab:participant_applications}
\end{table*}

\subsection{Promises of T2I in Medical Education \& Training}
\label{opportunities-training}
% \arpit{
% \textbf{F5} explained, ``\emph{We use Radiopaedia and use Google search when faced with challenging cases}''. \textbf{F2} mentioned how StatDx can be helpful when encountering really abnormal pathologies during training. 
An application context that was brought up frequently in the formative interviews, as well as our semi-structured interviews, was the potential to use this technology in medical education and training. R7 presents a compelling case for using text-to-CT Scan GenAI in medical education because, ``\textit{...it gives you the opportunity to identify and recognize something that is potentially super dangerous, that you then don't have to guess about when you see it in real life}''. 
% \begin{quote}``\textit{The great thing about simulation, and why simulation is being more and more incorporated into medical education at both the medical school level and at the resident level, is because it gives you the opportunity to identify and recognize something that is potentially super dangerous, that you then don't have to guess about when you see it in real life}''.
% \end{quote}
We see the essence of this quote throughout our conversations with students and trainees. We have grouped participants' insights on specific use cases into three themes regarding the promises of T2I GenAI in medical education and training: (\textbf{1}) supplementing traditional lectures and practice materials, (\textbf{2}) expediting learning through variations, differentials, and rare cases, and (\textbf{3}) empowering trainees during on-call or emergency situations.
% Radiologists frequently rely on Radiopaedia to support their own training and to educate peers. Their experience with Generative AI tools (such as Perplexity and ChatGPT) has been limited to general-purpose applications such as using them as alternatives to traditional search engines. The study revealed four specific opportunities for AI to provide value in radiology education. These are 
% }

\subsubsection{\textbf{Supplement Lectures \& Practice Materials.}}

Despite numerous resources out there, ``\textit{...there isn't a great place to be like, Hey med students, let's go here and learn chest CT today}'' (R8). R8 goes on further describing how they have never done anything in their classes where they get to interact with and compare the differences between abnormal and normal anatomies on CT scans. This could be due to how medical school lectures are designed, showing static images on lecture slides. R3 corroborates this limitation in existing lecture style by describing how they get content for their lecture slides by searching online or through their archives: ``\textit{I have this list of typical cases for all of these different lesions...choosing representative images from them...or showing a student the entirety of the scan}''. However, R3 went on to envision a scenario where they could use this T2I model to generate specific examples to use on their slides: ``\textit{I'm thinking about pre-populating...I'm gonna produce a lecture on this list of obscure topics...And I need specific examples of A, B, C, ... I feed those into your generator, and...I'll come back in an hour....and I'm ready to start producing my lecture.}'' R1 provides another perspective, acknowledging how convenient text-to-CT Scan GenAI could be when ``\textit{...teaching the residents or medical students,...we just want to show them something, we can just type it, and it will pop up}.'' 
% \begin{quote}
%     ``\textit{I have this list of typical cases for all of these different lesions, and I'm pretty efficient at going in and finding those cases and choosing representative images from them to display or showing a student the entirety of the scan}''.
% \end{quote}

% Text-to-CT Scan GenAI offers significant potential to enhance traditional educational methods from the learners' and the teachers' perspectives. From the teacher's perspective, 
% This scenario speaks to \textbf{Need 1}, fostering a design that enables exploration of user-defined pathologies between a student and teacher.
% \textbf{R3} expands upon that, detailing: ``\textit{The way this would be most useful is in rare diseases...ones that I don't have an example of, or that might be difficult to find on the Internet}''. 
% \textbf{R3} elaborated on a slightly different type of interaction with the GenAI for lecture preparation and its potential to address gaps in generating examples of rare diseases. They 
% This type of interaction similarly aligns with \textbf{Need 1}. 
% However, the interaction requires a slightly different design than the one described by \textbf{R1} as this is a passive interaction, whereas \textbf{R1} is more active and engaged.

From the students’ perspective, R7 and R8 saw the potential for text-to-CT Scan GenAI to support their studies from the earliest stages of their medical education. The process of medical students shadowing radiology residents is currently difficult as ``\textit{It's often been said that radiology even is very hard to shadow in because \textbf{it's like you're watching someone play a video game you don't know how to play}}'' (R8). R7 speaks to how this GenAI can ``\textit{lessen the learning curve}'' as students transition to trainees, exposing them to various pathologies before reviewing real patients. 
% recognizing how different lung cancers might look if it's invading vasculature versus if it's invading airway versus if it's central versus peripheral. So being able to recognize those different things so that you're not guessing when you're looking at it for the 1st time as a as a trainee. It can do a lot to lessen the learning curve. - R7
R8 reflected on how the tool could help medical students grasp foundational concepts and prepare for exams: 
``\textit{...being able to play with it...as you're doing questions, `Hey, here's what a pneumothorax looks like under tension, under no tension...I think would be invaluable.}''
% \begin{quote}``\textit{...a 1st-year medical student who knows nothing about anything with radiology, and is like `What is consolidation like? What is Cardiomegaly like?'...And then being able to play with it...as you're doing questions, `Hey, here's what a pneumothorax looks like under tension, under no tension...I think would be invaluable.}'' \end{quote}

Despite being able to similarly perform these interactions through Radiopaedia or StatDx, the students instead mentioned relying on other resources, such as UWorld or Amboss, for licensing exam prep. 

% This ability to visualize and interact with complex medical conditions could make learning more dynamic and engaging, something few learning resources provide them.

% ``\textit{If it can produce realistic examples of pathology, then we could use this as an alternative to looking something up on the Internet.}'' - R3

% [P?? for sessions where, like, you know, I'm reading throughout the day. But then, at the end of my day. I kind of like read off with my attendings, and if we're like going over something that can kind of look like something else, then, being able to pull this up and kind of like, show me on the fly, and and kind of teach you the the different features. Be super helpful.]

\subsubsection{\textbf{Expedite Learning through Variations, Differentials, and Rare Cases.}}

As medical students transition into their training program, it becomes increasingly important to get as much exposure as possible to a wide variety of cases. As R1 says, ``\textit{the more you see, the better radiologist you will become}''. However, despite relying on existing resources populated by the community and their training to see as many diverse images as possible, they are limited by what others \textit{can} share based on patient privacy regulations or what patient cases come through at their hospital. Radiology trainee, R2, highlighted how this GenAI could accelerate their learning by allowing them to study variations of diseases:
``\textit{...this would expedite that if I could just be like, `okay, I need to know the difference between this, this, and this'...I can just generate 5 different variances of those diseases on this tool with the different manifestations of those variants.}''

R2 exercises this during the interview session by prompting the model to generate `\texttt{lung cancer}', and then refines their prompt to generate a different variation of lung cancer. In their review of their generations, they put the two generated CT scans side-by-side. R5 similarly experiments with the granularity of the prompts by adding `\texttt{3cm}' to their initial prompt `\texttt{irregular speculated lesion...}'.
%, exercising \textbf{Need 2}. 

As many diseases can mimic each other on a CT scan, it is important to learn not only through the variations of how a disease can manifest but also through differentials of how the pathology of one disease is different from another. For example, R6 emphasized how this GenAI could make positive findings more accessible in a low-risk environment:
``\textit{Being able to see those positive findings is more educational than commenting on negative findings...sometimes we need to see, like...maybe more than 200 cases to see one positive finding...But if I can be able to generate those positive findings by just giving my impressions...I think that's something that can be very helpful.''}

Beyond variations and differentials, R8 imagined how this GenAI could be applied in day-to-day training scenarios to prepare for region-specific diseases. For example, they described a scenario where generated CT scans could be ``\textit{...included in my docket of cases I have to look at for the day and one or two of them might be training, region-specific training}''. They similarly mentioned the tool's value in training medical professionals in low-resource environments, such as being a ``\textit{...benefit to even Africa...And
being able to say, `Okay, here's what Cardiomegaly looks like with someone who might have tuberculosis.}'''

% These two scenarios highlight an additional design concept for medical text-to-image GenAI, which enables users to easily compare multiple prompt-generation pairs to each other for advanced learning in the viewing window. 
% \begin{quote} The more you see, the better radiologist you will become. So just seeing different variations of the same thing would be very helpful like if we could type one thing and then get different varieties of that thing in that respect it would be good for teaching - \textbf{R1}
% \end{quote}

\subsubsection{\textbf{Empowering Trainees when On-Call or in Emergency Situations.}}

Unlike clinical practice, where attending radiologists rely on their accumulated expertise, training is based on an apprenticeship model where trainees are closely supervised by their seniors. As trainees become more senior, they are increasingly put in situations where they may need to read cases in emergencies or on-call situations, where there is little to no time for doubt or waiting for an expert's approval. These situations demand quick and accurate decision-making, making tools like text-to-CT Scan GenAI an invaluable resource for trainees. Participants highlighted how this tool could serve as a critical support system during high-pressure scenarios. For instance, R2 described how it could aid in trauma cases by enabling them to create a checklist of generated CT scans based on diseases to look out for:
% , addressing \textbf{Need 2}:

\begin{quote}
``\textit{...when you're on trauma, you have a trauma come in...if I had this tool up and I had like 5 different generated pictures of the 5 things I'm checking for when a trauma case comes in...and then when the surgery team is there with me, `Hey, these are kind of like the things I'm looking out for', while we're searching through the image together}''.
\end{quote}
R2 expands this idea by imagining how they could use the tool to generate baseline CT scans to make sure they do not miss anything while on-call. R7 imagines how generating images for severe conditions ahead of shifts in the ER could accelerate trainees' decision-making, empowering them to handle emergency situations appropriately and promptly:
``\textit{...if you can generate your own images of what different aortic dissections look like, then, when you have a patient that comes into the ER with ripping chest pain, and you take them to CT, and you see, like, `Oh, sh*t! This is really bad, I need to call someone for help now', as opposed to `Okay, I haven't seen this before, but this sounds like what I think it could be like. Let me wait and see if I can get CT to read this, and then I have to call my attending', and that all adds time in an emergency situation...time is life.}''
R7 then prompted the model to generate an aortic dissection.

% , so that I can know, `Okay, I saw 5 cases that weren't real, but when they are real I'll know exactly what to do'.'' - \textbf{R8} \end{quote}

% \textbf{R7} went on to generate such a chest CT scan for aortic dissection, shown in Figure~\ref{fig:example-prompts2}, as they imagined in their scenario. 

% \begin{quote}
% being able to pull up this tool and generate kind of like like a baseline for for things we should be able to see, you know, kind of helps like, okay, like. I'm going through a bunch of like studies on call. And I need to make sure I don't miss any like any like
% massive, like tumor, or, you know, like a lung hematoma or something like that -p2
% \end{quote}

\subsection{Promises of T2I in Clinical Practice}
\label{opportunities-practice}

% \arpit{
While GenAI can have a high potential for medical education and training, it remains to be seen whether and how it can augment clinical workflows. Clinical practice is slightly different from training because during training, the junior trainees, ``\textit{...are reading every case with the attending}'' (R6). 
% In clinical practice, as a board-certified radiologist, you are signing off on cases yourself. 
Our study further addresses \textit{RQ1} by identifying how text-to-CT Scan GenAI can augment attending radiologists in clinical practice, including (\textbf{1}) improving report impressions of medical images, (\textbf{2}) planning and communicating with others, and (\textbf{3}) supplementing the existing resources they rely on when faced with challenging cases. 
% }

\begin{table*}
    \centering
    \begin{tabular}{|c|l|l|}
        \hline
        \textbf{Category} & \textbf{Challenges} & \textbf{Participants} \\ 
        \hline
        Technical Challenges & Generated CT scan resolution quality  & R1, R3, R6, R7, R9\\ 
        & Transparency and Hallucinations & R3, R4, R9\\ 
          & Contrast-Enhanced CTs & R3, R4, R9 \\
           & Generation Expectations & R3--R5, R9 \\
          % & Generation time & R3, R9 \\
          \hline
        Domain-Specific Risks
         % & Prompting Guidance & R1, R4, R6, R7 \\
         & Confirmation Bias \& Distilling Incorrect Information & R2, R3, R6, R7, R9 \\ 
         & Over-exposure to synthetic CT scans during training & R3, R5, R7 \\
         
        \hline
    \end{tabular}
    \caption{\textit{RQ2}: Participants mapped to which technical challenges and domain-specific risks that they brought up.}
\label{tab:participant_challenges}
\end{table*}

\subsubsection{\textbf{Improving Report Impression of Medical Images.}}

Interpreting medical images to identify disease progression or abnormalities often depends on differential diagnosis and validating speculations. Participants highlighted how text-to-CT Scan GenAI could play a pivotal role in supporting these tasks by generating CT scans based on their interpretation to visualize hypotheses and refine interpretations. For example, R1 described how the GenAI could assist in determining conditions like cardiomegaly by providing reference images to compare against patient scans.
% , addressing \textbf{Need 2}.
% : ``\textit{Is it a cardiomegaly, or is it not a cardiomegaly?...that is also very useful to see if there is a cardiomegaly or if it's just because of the patient's positioning}''. 
Beyond aiding in diagnosis, the tool could support the refinement of more precise clinical impressions, which participants discovered as they continued generating more CT Scans. R5 described how the generation of reference images could refine the specificity of their observations, ``\textit{Instead of just right pleural effusion, I can compare the image and then specify which segment I expect the pleural effusion to be in}''. Using this process, R5 reflected on how generating CT scans ``...\textit{can be useful to revise my diagnosis to add more details''} and ``\textit{...use the generation for differential diagnosis}''.
% \textbf{R5} also noted how generating CT scans could help them clarify ambiguous findings: ``\textit{Sometimes, pus in the pleural cavity can misrepresent as soft tissue density, so if I am curious, I can use the generation for differential diagnosis}''.

The comparative functionality in the interface allowed R6 to discover that their impression of the original patient case presented to them was not completely correct when they compared the image findings in the generated image to the provided real CT: 
``\textit{So, I said, smaller effusion, and sometimes people say mild effusion, some say severe effusion. But so this generated AI gives me a very small effusion. But \textbf{now I'm thinking this might be a mild effusion instead of a small effusion}, which can be very helpful actually...Something like when we use those subjective words and if we have something where somethings objective, could help us a lot actually.}''
% \end{quote}  

% This type of scenario presents a true clinical scenario where a radiologist or senior trainee can effectively use this GenAI when faced with a patient case they need to read and may have doubts about the severity of the disease or which disease is present. 
Together, these scenarios underscore the potential of text-to-CT Scan GenAI to enhance the accuracy, efficiency, and precision of image interpretation in clinical practice, empowering clinicians while reviewing both routine and challenging cases. However, it also raises implications for accountability, confirmation bias, and unintentional deskilling.

\subsubsection{\textbf{Planning \& Communicating with Others.}}

An interesting use case outside of direct diagnosis or education that was mentioned was in planning and communicating with other stakeholders. For example, R2 described how generating CT scans could be ``\textit{...\textbf{super helpful in surgical planning}...you can show the patient like, `hey, this is going to be what you look like post-surgery.}'' In a similar vein, R8 describes how they could imagine a scenario where they show a patient how their condition looks compared to a normal condition.
% [“That we're kind of using these models to kind of see how a disease presents in your body so that we can plan ahead.” - R2]

Radiologists are not the only ones who will be interacting with medical imaging throughout the clinical workflow. This could introduce multiple impressions that may disagree about a diagnosis. As such, R5 brings up how ``\textit{\textbf{These generated images can act as additional evidence when I am communicating my diagnosis to other doctors}}.'' 
% \textbf{R7} expands upon this idea by emphasizing the importance of surgeons being able to read medical images because they ``\textit{really get an understanding of what we should expect before we go in}''. 
R7 motivates R5's point, noting how some surgeons they have shadowed will ``...\textit{trust their own reading of the image over whatever the official [radiologist's] read is}.'' This is an interesting dynamic in the clinical workflow where generated CT scans could aid in these collaborative workflows, helping radiologists provide validations of their reports by generating CTs that show differentials for other physicians to read.
% , then they just say, like, Okay, this is what I think it is. And you know, I might disagree with the radiologist. And this is something that we can talk about further}''.\end{quote}

\subsubsection{\textbf{Visual Memory Support.}}

Radiologists are constantly learning beyond their formal training programs, often needing to recall visual features of pathologies they may not have encountered recently. As R4 describes it, ``\emph{I think the more it passes from graduation, yeah, the more you forget.}'' As a result, participants highlighted how text-to-CT Scan GenAI could serve them ``\textit{by strengthening our visual memory}'' (R4). This is particularly valuable for general radiologists working across different specialties where they encounter CT scans of the chest along with other regions of the body. As R5 explained:
``\emph{I’m working in a different field of radiology. If I am specialized in one branch, and I see a different CT scan, this can be useful to remember something you haven’t seen in a long time.}'' R5 goes on to say ``\textit{It would be easier to use this tool instead of Radiopaedia when I was a student, I would just type my prompt and see an image much faster than having to search for the right case}. R3 and R4 similarly conclude this could be used as ``\textit{...an alternative to looking something up on the Internet.}'' 

\subsection{Risks \& Challenges with Medical T2I GenAI}
\label{challenges}
% \arpit{
As a result of key stakeholders generating CT Scans for pathologies beyond what was shown to be validated by~\citet{xu2024medsyn}, we discovered technical challenges and domain-specific risks, addressing \textit{RQ2}. This section is separated into two parts: (\textbf{1}) technical challenges and (\textbf{2}) domain-specific risks, each with sub-themes shown in Table\textbf{~\ref{tab:participant_challenges}}. These groupings were identified from a thematic analysis of quotes related to model behavior and implications of model use in practice. For the interpretation of our findings, technical challenges refer to model development, and domain-specific risks refer to downstream impacts from model use. Such risks and challenges may act as barriers to the adoption of medical T2I GenAI if left unaddressed.
% }
% \begin{figure*}[!h]
%     \centering
%     \includegraphics[width=1\linewidth]{figures/Challenges_Table.pdf}
%     \caption{\textbf{Sample of Participants' Prompts \& Generated Outputs to Discover Technical Challenges.} \textbf{R3} chose this prompt to probe whether the model could actually generate a contrast-enhanced CT, which, as shown, cannot. \textbf{R7} chose this prompt to test multiple pathologies, more complicated pathologies, and jargon/acronyms. }
%     \label{fig:example-prompts2}
% \end{figure*}

\subsubsection{\textbf{Technical Challenges.}}

% [Talk about the need for minimal hallucinations and higher-resolution imaging] [talk about the need to generate things quickly despite these models being computationally expensive] [ the need to come up with transparency methods that go from text to 3D output]
Impressively, the model was able to generate pathologies that were not tested in the original study~\cite {xu2024medsyn}. However, there are four primary technical and AI safety challenges that arose while participants were generating and exploring CT scans: (\textbf{1}) the resolution quality of the CT Scans, (\textbf{2}) the approach to validation and hallucinations, (\textbf{3}) generating contrast-enhanced CTs, and (\textbf{4}) technical expectations of the model generation. 

Five participants commented on the resolution of the generated CT scans being of lower quality, even though the model we used was designed to develop high-resolution CTs. While looking through their generated CT scan, R7 found it difficult to see if the pathology they asked for was there ``\textit{because of the resolution}''. They described the experience as if they were looking at a CT scan on a really bad computer instead of the expensive computers that the radiologists use, while R9 described it looking ``\textit{like they moved during}'' the CT scan. R6 similarly asked during their session if they had a bad internet connection or if it was just the generation quality. However, R3 clarified that a radiologist may not necessarily require a high-resolution CT scan as long as it offers the required information to support the goal of the generation: 
``\textit{It's not necessary to fool me into thinking it's a real CT. What you want it to do is \textbf{provide a sufficient example that I would recognize this on a real CT}}.'' 
% “comparing this to a real image is like this is what a Ct will look like on like a really shitty computer like, it's a good Ct, but it's on a shitty computer. So the resolution is low compared to like. If you go to the reading room and you're looking at it on the same. The same. Read on the the radiologist. Super expensive computers. It'll be a lot more sharper. So this, to me, is like a Ct scan on a bad computer.” - R7

% Second, one of the most important challenges with introducing new technology like this is validation to gain wide acceptance. \textbf{R7} points out that validation would need to ``\textit{...make sure that this is something that's not going to hurt patients, and indirectly, [give] poor quality education.}'' This raises an interesting challenge into what success metrics look like for a tool like this; for example, minimizing hallucinations and maximizing anatomically accurate generations (quantitative metrics) or maximizing high-quality education and report accuracy (applied metrics) could be difficult metrics to measure and are beyond the traditional human-AI collaboration team performance metrics. 
Second, several participants encountered flaws in the generated output as they were prompting the model on image findings that were not tested for by the quantitative model evaluations. For example, R3 spent time closely auditing the outputs, commenting that the model has ``\textit{...thrown all sorts of pathology onto this patient...}'' including ``\textit{...metastatic bone disease...and a plural recess down here that doesn't connect to anything}.'' R4 and R9 similarly identified missing or injected pathologies in their generations. A vast majority of the ML literature refers to this type of model behavior as hallucinations, which come with domain-specific risks. As with any black box AI, it is important that users can see which part of the prompts are associated with which regions of the generation. This becomes especially critical and challenging with a 3D-generated output where multiple pathologies are combined within a single prompt. This task, however, is computationally expensive and underexplored for text-to-image GenAI~\cite{bradshaw2025toward}.
% \textbf{R8} raises the point that having a really good model with few errors could be more dangerous than a half-decent model because it may be harder to appropriately trust a really good model compared to a half-decent model.

Third, R3's prompt (``\texttt{Necrotizing Pneumonia on
contrast-enhanced CT}'') asked for a contrast-enhanced CT scan, which is a specific type of CT scan that shows reactions of a fluid given to a patient before the CT scan~\cite{mayo_ct_scan}, to probe the model's limitations further. As discovered, the model currently cannot produce contrast-enhanced CT scans, as it was not trained on contrast-enhanced CT scans. R4 and R9 also acknowledged this limitation when they generated something that they expected to be shown on a contrast-enhanced CT. Without discussing with radiologists the importance of contrast-enhanced CTs in accurately determining certain image findings, models may be trained on data that does not represent what radiologists would benefit from. Similar to this, R1 and R6 mention how report impressions can sometimes be subjective (\textit{i.e.}, how severe a disease actually is), which is especially consequential in training models depending on whether the reports are sourced from trainees or experts. R9 mentioned how they struggled to get a certain pathology to be of less severity, and they concluded that ``\textit{maybe all the patients you guys had were more like outpatient.}''

Lastly, despite the GenAI's potential value, R3 and R9 acknowledge that the generation time delay is substantial ($~3$ minutes), especially if comparing it to searching for examples on Radiopaedia. R9 said that as residents, they have ``\textit{...limited time while you're reading images to do research.}'' Beyond generation time, several participants commented on the GenAI's limitations compared to content provided by existing resources, such as Radiopaedia. For example, R5 contrasts the model \& interface to how ``\textit{Radiopaedia also has some additional notes [about the image finding] which can be helpful for differential diagnosis}.'' These resources allow them to parse detailed descriptions, aiding in their conclusion about a diagnosis, whereas the GenAI is limited to taking a single prompt at a time to generate a CT scan and does not provide any detailed description in return.

\subsubsection{\textbf{Domain-Specific Risks}} 
The development of medical T2I that addresses stakeholders' needs and concerns goes beyond solving technical challenges. We must also consider the complex interplay between user expectations and domain-specific implications. Our findings revealed two domain-specific risks that need to be considered to ensure that T2I GenAI meets the needs of the key stakeholders and aligns with complex workflows: (\textbf{1}) mitigating confirmation bias and the distillation of incorrect information and (\textbf{2}) supporting continuous exposure to real medical images.

% Expectedly, many participants had their own unique challenges with coming up with a prompt. R1, R4, and R5 all experienced uncertainty around how detailed their prompt could be, which turned into iterative prompting, making small updates to the prompt to wait another three to four minutes to see if their new generation is closer to what they were expecting. R7 was unsure whether the prompt could handle jargon and acronyms. This speaks to the need for guidance on how prompts can be constructed based on how the model was trained.

First, participants were unsure if and where the requested pathology from the prompt was generated (R2, R6, R7, R9). While reviewing the generations, they would say something like ``\textit{Maybe this is what they [AI] wanted}'' (R2) or ``\textit{I potentially see that dissection}'' (R7). This type of behavior with AI, masked as confirmation bias, can be detrimental to a trainee's learning progress. R9 highlighted potential consequences: ``\textit{If you're learning and then you get like false information, uh oh, you know, that's confusing. \textbf{And you don't know if you're wrong or [the] software is wrong}}.'' In the use case of practice exams, these unprompted pathologies could distract more than support the student (R3). Ultimately, it is possible for students generating CT scans who are unaware of the additional, missing, or misrepresented pathologies to unintentionally learn the wrong things. For example, after R7 generated a CT scan for aortic dissection, they describe a potential issue related to confirmation bias: ``\textit{\textbf{Am I seeing what I'm actually seeing? Or am I seeing what I want to see?}}'' To this point, R7 emphasizes that research would need to ``\textit{...make sure that this is something that's not going to...indirectly, [give] poor quality education.}''

% Nonetheless, R3 described how this type of transparency would actually be a great educational component as it now can direct learners to exactly where a particular disease pathology is manifesting.
% \subsubsection{\textbf{Domain Design Challenges}}

% [talk about the need for additional context around the case that was generated, similar to case context that radiopaedia provides]
Second, despite the participants acknowledging that the generated CT scans were of lower resolution, they were surprised by the fact that they were looking at a generated CT scan. R7 remarked, ``\textit{\textbf{If you hadn't told me I wouldn't have known [that the CT was AI generated]}}.'' F2 reacted similarly, saying ``\textit{Oh wow that’s really good because I couldn’t tell which one is from which}.'' R5 stated ``\textit{With the right prompt, the quality of the output is so good that I cannot distinguish that this is AI, if I did not create the prompt myself}.'' This in itself raises potentially consequential implications as R5 states that ``\textit{Generated images can be useful for experienced doctors who have seen thousands of images, but \textbf{residents need to develop experience with real patient cases and images}}.'' From a lecturer's perspective, the students want ``\textit{...to see real examples with validated pathology}'' (R3). This underscores the ongoing theme that radiologists and those in the broader medical field need to continue seeing a plethora of real CT scans throughout their careers, given that the ones they need to interpret are real patients. 

% Beyond the importance of gaining exposure to real CT scans, 

%%% ------------- DISCUSSION ---------- %%%%%

\section{Discussion}
\label{discussion}

Building advanced models, such as the text-to-CT Scan GenAI, often occurs without a deep understanding of key stakeholders' needs and potential safety risks. This lack of clarity raises important questions about misuse, misinformation, and the potential erosion of expert skills, particularly in high-stakes environments such as medicine~\cite{karabacak2023advent,ooi2023potential}. 
% Our work builds on and formalizes the growing body of research highlighting the potential of text-to-image GenAI beyond traditional applications. Our work also extends ongoing RAI research by identifying technical challenges and domain-specific risks of generating synthetic medical images.
Below, we discuss the benefits of our study design and the implications associated with the use of text-to-image GenAI for radiology

\subsection{Evaluating Domain-Specific Text-to-Image GenAI}
% \textbf{Value in using Generative AI-based Design Probes as a Research Lens.} 
% Without being able to interact with the GenAI, we saw our formative study participants greatly struggle to identify the concrete value that text-to-CT Scan GenAI could bring to their workflow. Within radiology, traditional AI applications, as well as text-to-text or image-to-text, are more prevalent, making it easier to imagine and discuss. We similarly observed this with the design probe participants before they actually interacted with the probe. However, after generating a CT scan themselves and interacting with the output, they were able to easily realize its potential across several scenarios. Addressing \textit{RQ1}, this is an interesting takeaway for design researchers, showcasing the power that generative AI-based design probes have in researching design opportunities and challenges of novel GenAI applications with domain experts who may have little knowledge or experience with that technology. Using a Research through Design approach allowed us to elicit rich information from domain experts about a very novel technology.~\citet{epstein2022happy} similarly leverage generations as a medium to elicit ideas from participants.

\textbf{Human-Centered Evaluations of T2I GenAI with Stakeholders.} 
As an outcome of conducting exploratory model prompting with key stakeholders, we observed how each participant in our study was interested in exploring and evaluating a different capability of the model. Unlike the formative interviews, where users could not prompt the model, the exploratory model prompting study design allowed us to uncover numerous technical challenges and domain-specific risks. For example, it helped expose that the thumbs-up/down feedback mechanisms offered for each generated image in the interface did not address what participants really wanted to audit, which was the ability to generate contrast-enhanced CT scans, the resolution of the images (similar to~\citet{brisco2023exploring}'s observation), or the misrepresentation of image findings. 

We also learned that CT scan impressions can be subjective, making it important to be concerned about the expertise of the radiologists who wrote the reports for the training data. Using reports by trainees or unverified by multiple radiologists could introduce inconsistencies that can impact the reliability of the model. All of these challenges are difficult to quantitatively capture; however, conducting exploratory human-centered AI evaluations to simultaneously elicit this will be useful to the advancement of RAI.

Lastly, unlike many existing risk taxonomies and evaluations for T2I models, we observe a specific set of risks that the radiologists were primarily concerned about: distilling incorrect information and losing exposure to real CT scans during training. These risks represent unintentional harms that could come about from this technology being widely available in practice. It is interesting that more granular risks or harms, such as various forms of bias, were not acknowledged by the participants. Despite this, it will be critical to identify domain-specific biases that arise with domain-specific T2I models and innovate solutions to mitigate them.
% Some stakeholders mentioned the value of being able to provide region-specific training using this model. This introduces the importance of working with stakeholders to audit the model's capability for region-specific diseases. ~\citet{koohi2023generative} and~\citet{currie2024gender} speak to the importance of mitigating gender and ethnic bias across image datasets to minimize the generation of inaccurate images.~\citet{yildirim2024multimodal} found that radiologists value in having a local-patient population-based experience, speaking to the importance of developing these models closely with key stakeholders.

\subsection{Implications of Text-to-CT Scan GenAI in Medical Education, Training, \& Practice}

% \textbf{Designing for Variations \& Differentials.} 
We allude to \textit{RQ1} as we discuss the implications of T2I for medical education, training, and practice. Our participants emphasized the role that online resources and pure experience play when faced with challenging cases and professional growth, as similarly observed by~\citet{yildirim2024multimodal} and~\citet{rassie2017apprenticeship}. \citet{ElAli2019} and ~\citet{Quaia2023} have observed an increase in students' and trainees' performance when provided access to radiology-specific repositories, emphasizing the importance of supplemental resources to traditional lecture materials. Extending these insights, five participants referenced how getting exposure to variations can play an important role in competency within medical imaging-based professions. Generating variations as a design has similarly been explored by~\cite{gero2024supporting}, highlighting the value that variation plays in sensemaking. However, the design of our GenAI plugin did not easily enable the generation and visualization of variations for a single prompt. This was also difficult due to the generation time for a single CT scan. As~\citet{weisz2024design} suggest, it is important to design GenAI applications to enable variability and expose differences from a prompt. While this generates variability aimed at meeting the user's needs, we also see how this can expedite education by showing several CT scans with the same pathology in several variations. This would help prepare students to practice identifying serious conditions before dealing with real patients, a benefit that~\citet{munuera2023generative} similarly mention. 

Our participants also stressed the importance of being able to understand what real human anatomy is like by looking at thousands of real CT scans. This opens the door for RAI work to explore various design opportunities for domain-specific T2I models, such as inpainting capabilities. Furthermore, RAI can explore the human-centered needs around explicitly identifying which case in the viewing window is generated and which is real to avoid misuse and abuse of the GenAI capabilities~\cite{munuera2023generative}.

\textbf{Considering Collaborative Workflows.} Our participants highlighted the collaborative workflows that take place within medicine and how text-to-CT Scan GenAI could provide radiologists an opportunity to provide generations as rationalizations for their reading to enhance their collaboration with surgeons.~\citet{yildirim2024multimodal} also saw the value of showing similar images as a use case for reassurance between radiologists and other medical professionals. Such an interaction throughout collaborative workflows might require designs that allow radiologists to leave additional notes or engage in an asynchronous discussion with the surgeon about their generations and patient case impressions. 
% Case overload for residents and attendings caused by improper integration of these tools.
% [talk about the potential: won’t these cases bring more work up to the residents or the attendings? Make sure this doesn’t unintentionally impact case overload already for the residents and attendings]

\subsection{Safety Implications of Medical T2I GenAI}
% \textbf{Designing to Empower Medical Professionals instead of Overwhelm them} \color{blue}\textbf{NEED TO PUT TOGETHER}\color{black}
With any powerful text-to-image GenAI model, it is necessary to consider the broader ethical and safety implications. We allude to \textit{RQ2} while discussing design implications for safeguarding against hallucinations and confirmation bias, as well as addressing misuse of the generated CT scans. 

\textbf{Safeguard Against Hallucinations \& Confirmation Bias.} As we have seen through our own experiences and previous work, hallucinations are prevalent in GenAI systems. This is especially concerning because primarily the more experienced participants were commenting on misrepresentations of the prompt in their generated CT scan. This motivates future works to explore ways to express hallucinations to users, especially to novice users. Along with hallucinations, some participants raised concerns about confirmation bias, which can emerge when working with AI~\cite{rosbach2025two}, especially among students and junior trainees. While it may be hard to avoid generating hallucinations, we can design mechanisms to safeguard end-users from confirmation bias during their learning experience. Designing mechanisms to minimize confirmation bias is especially important in an educational environment because recent work has shown how people can start reasoning incorrectly when given misleading explanations generated by AI~\cite{spitzer2024don}. These designs are especially important to avoid deskilling, slowing professional growth, and minimizing over-reliance. One potential solution that we see is by developing visual explanations that associate concepts from the prompt to regions of the generated image, such as the technical solutions proposed by~\citet{chefer2023attend},~\citet{bradshaw2025toward}, and~\citet{evirgen2024text}. 
% Alternatively, another design to address this is by leveraging the existing infrastructure for image-to-text GenAI and presenting a complete pipeline, comparing the user's prompt to the generated text. However, this approach can surely introduce more hallucinations than addressing the problem.

\textbf{Preventing Misuse and Malpractice.} Although this was not an emerging theme from the interviews, it is necessary to anticipate potential misuse and malpractice of synthetic medical images.~\citet{koohi2023generative} suggests the potential for AI-generated medical images to fabricate clinical results or patients' conditions, which can lead to inaccurate conclusions or mistreating patients. 
% Expanding upon this, insurance agencies or lawyers could wrongly use these to refuse treatment or sue medical providers for mistreating patients. 
The misuse and malpractice scenarios could go on, but it is more important to discuss practices to prevent such scenarios from becoming prevalent. This especially becomes important when these models increase competency by being able to do tasks such as inpainting original patient CT scans.~\citet {horvitz2022horizon} proposes watermarking as a potential solution to distinguish synthetic from real content, promoting transparency. 
% Policymakers and AI regulatory bodies should be involved throughout the entirety of text-to-CT Scan GenAI design and deployment pipelines.

% \textbf{Enabling Seamless User-Defined Pathology Generation \& Visual Comparisons.} \color{blue}\textbf{NEED TO PUT TOGETHER}\color{black}
% ["images found online were often poor quality and did
% not reflect the local patient population...Future work
% should explore how VLM capabilities lend themselves to unique
% search, filter and retrieval experiences that go beyond what web
% search can provide"~\cite{yildirim2024multimodal}]
% [designing interfaces to have guidance on the prompt expectations and constructions] - [interface should set expectations of the interactions that are going to be possible with the model] 

% [design from ~\citet{yildirim2024multimodal} focused on comparative queries as well because not only is it helpful for differential diagnoses but also helpful in see patient's progression - this is also a key design concept that would facilitate many of the applications that we learned about from the participants. Simiarly,~\cite{yang2020xraygan} present a comparative pipeline for medical text-to-image models.]

\section{Limitations \& Future Work}
\label{limitations}
This work comes with limitations that future work can build upon. First, our study involved a small sample size for both the formative study and semi-structured sessions. However, this sample size choice is restricted by domain stakeholders' availability and respect for their time. Additionally, our participants all had different levels of experience and perspectives about AI, which could have tainted their perspectives on the promises, risks, and challenges that they identified. While we had briefly asked participants about their prior experiences with AI in the pre-probe sessions, we have not done an extensive analysis of that data. Also, our study is limited to a very domain-specific T2I for chest CT scans, which could potentially limit how our conclusions translate to other medical text-to-image GenAI. Although not all participants were chest specialists, we felt that they were still able to successfully identify valid promises, risks, and challenges even beyond chest CTs. For example, two participants were able to see how this text-to-image generation process could generalize to other imaging modalities. % (\textbf{P1}, \textbf{R8}). 

Additionally, we did not conduct a detailed analysis of all the prompts generated by participants, as this was not our primary goal with this study. Furthermore, we have a limited number of prompts/generations from participants, as each generation took between three and four minutes. We did not intend to further evaluate the model's performance or identify heuristics for prompts that work better with this model for this study. These areas remain for future exploration to better understand how key stakeholders would prompt generative models depending on their goals.

Similarly, we did not perform quantitative evaluations of the accuracy of the generated CT scans from the radiologists’ perspective during these sessions. Future work should consider including quantitative model auditing activities with stakeholders using familiar medical interfaces and tools to maintain external validity and to ensure that evaluations appropriately target stakeholders' concerns. 
% Conducting these audits and evaluations within familiar medical imaging interfaces, rather than generic survey platforms, is critical for maintaining external validity.
% , as radiologists rely on specialized functionalities not typically available.

% Another important direction for future work is the development of more specific design concepts. In this study, our generative AI plugin primarily served as a design probe to provoke discussions and gather insights. Thus, we did not go in with any formal design concepts specific to each application context. Future work should explore how interface designs change based on their use in education, training, and practice. Doing so would help facilitate richer, more actionable design feedback from stakeholders.

\section{Conclusion}

Medical text-to-image GenAI capabilities are increasingly being investigated as models become increasingly competent. Given the lack of human-centered research involving key stakeholders on the promises, risks, and challenges of medical text-to-image GenAI, our work seeks to bridge this gap by conducting exploratory model prompting sessions with a text-to-CT Scan generative model. Our study highlights the opportunities of text-to-image GenAI in medical education, training, and practice, such as empowering students to learn in low-stakes environments and augmenting radiologists when faced with challenging cases. These promising opportunities expose technical challenges and domain-specific risks of domain-specific T2I models that the RAI community can consider, such as confirmation bias. Overall, our work represents the importance of evaluating T2I models from a human-centered lens, taking domain-specific needs and challenges into consideration. We hope this work encourages RAI researchers to further explore the risks and challenges of domain-specific T2I models.

% \input{appendix}


% \section{Positionality Statement}
% Two authors are considered as domain stakeholders, one as a board-certified radiologist and the other as a resident (trainee) in a radiology program. A portion of the author team consists of machine learning researchers and the another portion of the author team consists of human-centered AI researchers.

\section{Acknowledgments}
Thank you to Dr. Ahmed El-Sherief and Dr. Samir Shah for insightful informal conversations about the implications of our work. Thank you to everyone who helped us recruit participants. Thank you to Wesley Deng for providing feedback on the draft manuscript. ChatGPT4, Writeful, and Grammarly were utilized throughout the sections of this paper to improve the authors' writing, such as grammar, clarity, flow, and spelling. 

\bibliography{aaai25}

\section{Supplemental Material}

\section{Formative Study}
\label{formative}
Designing human-centered generative AI systems in high-stakes domains can be difficult, especially when the applications of the technology are under-explored. We conducted short formative discussions to lay the groundwork for our design probe by capturing key stakeholders' perspectives on the potential value of text-to-CT Scan GenAI. These sessions were not intended to be comprehensive but to serve as an initial step to guide our design probe development. We describe the methods used, followed by the resulting needs of stakeholders.

We conducted IRB-approved formative interviews with seven stakeholders to explore their needs. %Given the exploratory nature of this stage, and the busy schedule of our medical stakeholders, we  kept it short (around $10$ minutes) 
The sessions were designed %to last $10$ minutes in order 
to gather an initial understanding of radiologists' current perceptions and expectations of medical text-to-CT Scan GenAI. Although previous works suggest using medical text-to-image GenAI in medical education and clinical practice, we used these interviews to help capture any additional contexts throughout radiology where text-to-CT Scan GenAI could be useful to inform our design probe. Lastly, these sessions helped us finalize the different groups of participants to recruit for the design probe, such as medical students and junior trainees (who we initially did not consider stakeholders but found to play a critical role in the process).

Before the %$10$-minute 
interview session, the radiologists were shown a series of pre-generated CT scans. Similar to~\citet{waikel2023generative}, the radiologists were blinded to the exact prompts used to generate the CT scans. During the recorded interview session, they were told that the CT scans were generated and were asked questions about (\textbf{1}) their perceptions of text-to-image GenAI being used in radiology and (\textbf{2}) the potential areas in their workflows where they think text-to-CT Scan GenAI could be of value. These sessions were designed to be short (around $10$ minutes) because radiologists' time is valuable, and we did not intend to use this to comprehensively capture rich data for our research questions. The $10$ minutes was enough time to guide us through building a contextually accurate and needs-informed design probe. The interviews were recorded and transcribed to synthesize findings.

\subsection{Participant Recruitment} Participants were recruited through our clinical collaborators' network. For each session, we provided the participants with an overview of the interview's purpose and obtained their consent for participation. All participants consented to audio recording for data analysis and were provided a form of monetary compensation for their time. We interviewed six attending radiologists and one senior radiology trainee (\textbf{F1--F7}) across two institutions; demographics are shown in Table\textbf{~\ref{tab:phase1_demographics}}. Sessions were held on Zoom or Microsoft Teams, based on the participant’s preference, and lasted for an average of $6.8$ minutes.

\subsection{Analysis}
% \arpit{
We used an affinity diagramming approach to analyze the qualitative data. Ideas from the transcripts were pulled out as quotes, and affinity mapping was conducted to group similar ideas across interviews based on common themes. Multiple authors sequentially reviewed the groupings individually, and all the quotes were grouped into primary themes. Sub-themes were collectively generated and agreed upon.

\begin{table}[!t]
    \centering
    \begin{tabular}{ccc}
        PID & Expertise Level & Experience (years)\\
        \hline
        F1  & Expert & $\geq$20 \\
        F2  & Senior Trainee & 4 \\
        F3  & Expert & 5--10 \\
        F4  & Expert & 5--10\\
        F5  & Expert & 5--10\\
        F6  & Expert & 5--10 \\
        F7  & Expert &  $\geq$20\\
        \hline
    \end{tabular}
    \caption{Participant Demographics for the formative study.}
    \label{tab:phase1_demographics}
\end{table}
 
\section{Understanding Perceptions, Needs, \& Workflows}
\label{needs}

As a result of our formative discussions, we identify training (after medical school but before becoming an attending radiologist) as an additional application context beyond medical education and clinical practice. Overall, we identified three needs across education, training, and practice to guide the development of our design probe. 

% \subsubsection{\textbf{Need 1: A More Targeted Approach to Explore User-Defined Pathologies}}
\subsection{Need 1: Visualize the Exact Pathology Searched for to Enhance Training}
We learned from our formative discussions that radiologists rely heavily on a combination of community-driven and paid resources, such as StatDx~\cite{StatDx}, Google~\cite{Google}, and Radiopaedia~\cite{Radiopaedia}, to explore different diseases and confirm pathology descriptions when encountering challenging or unfamiliar cases (\textbf{F1}, \textbf{F4}--\textbf{F6}). Radiopaedia is an online community-driven resource for trainees and radiologists to share and discover patient cases, imaging, and disease pathologies. StatDx is an encyclopedia for different pathologies. However, these tools can require time-intensive searches, and even then, they may not provide exactly what is needed for thorough differential diagnoses. For instance, \textbf{F2}, a senior trainee, described their current process when reviewing challenging cases:

\begin{quote}
``\emph{If this is something that is really abnormal and I don’t know, we have some kind of library called StatDx. All residents have that. So \textbf{I usually look the disease up...but you have to search through it}. It’s kind of useful sometimes. You google, for example, pulmonary edema, then it gives you a description and the findings in different modalities and images...But most often we use Google and Radiopaedia...}''.
\end{quote}

Even with these resources, trainees may still struggle to find exactly what they are looking for due to the lack of publicly accessible reference images. These community-driven or paid resources are ultimately limited by what others \textit{can} publicly share based on patient privacy regulations, making it challenging to explore complex pathologies. As \textbf{F2} points out: ``\emph{...\textbf{there are some rare cases that you cannot see the actual images and you have to learn it from the textbook}...}''.
These limitations highlight the need for a tool that enables visualization and exploration of specific pathologies. Text-to-CT Scan GenAI offers a powerful solution to fill this gap by allowing trainees to generate CT scans of the exact combination of pathologies they are searching for. Beyond training, \textbf{F2} and \textbf{F5} agree that this GenAI has potential in teaching and education. Acknowledging the limitations of existing resources, \textbf{F6} said 
\begin{quote}``\emph{It [text-to-CT Scan GenAI] helps us for learning about rare diseases that maybe they have not seen during residency or training curriculum...AI can give them [trainees] a chance to look at rare and interesting cases}''. 
\end{quote} 

% , helping them ``\textit{increase the certainty of our diagnosis}''. 
These observations reveal that radiologists are already well accustomed to searching for pathologies on community-driven and paid resources to retrieve CT scans, albeit they can only, at best, find images \textit{most similar} to what they searched for. Text-to-CT Scan GenAI can address this by providing the ability to create CT scans that exactly match the pathology searched for and handle combinations of pathologies in searches that are not publicly accessible. Additionally, the educational potential underscored by these findings suggests the importance of seeking input from junior trainees and medical students to further explore how this technology can enhance learning and training experiences.

\subsection{Need 2: %Seamless 
Integration of Visual Comparative Analyses in Clinical Practice}
Radiologists beyond training often rely on online resources, such as Google and Radiopaedia, to address challenging cases (\textbf{F1}, \textbf{F4}--\textbf{F6}). These tools are helpful in obtaining pathology descriptions and imaging in multiple modalities. However, they fall short when radiologists need to compare multiple CT scans for differential diagnoses or to validate their impressions of a case. As \textbf{F6} described: ``\textit{Radiopaedia gives you some structural explanation about a patient below the image and you can find related images}''. \textbf{F6} imagined how text-to-CT Scan GenAI could enable a more interactive process while reviewing scans and writing up reports:
\begin{quote}
''\textit{Maybe we use AI to generate from the radiologist report a new CT scan and \textbf{compare this CT to the real one and understand our errors} and mistakes and which part may be ignored or may be misdiagnosed and help us to correct our reports.}'' 
\end{quote}

This underscores the need for tools that not only provide reference images but also facilitate side-by-side comparisons of multiple CT scans, allowing radiologists to uncover overlooked details and validate their findings. Similarly, \textbf{F1} described how AI could help radiologists identify pathologies more effectively, while \textbf{F4} highlighted the potential of AI to assist in double-checking their impressions by helping them ``\textit{...find some clues that maybe our eyes lose when we want to put all the findings together to make a proper diagnosis}''. 

These perspectives suggest that text-to-CT Scan GenAI embedded in existing CT scan viewing applications could augment radiologists' clinical practice by enabling more efficient comparisons, uncovering overlooked details, and supporting decision-making in critical scenarios.

\subsection{Need 3: Building Familiarity with text-to-CT Scan GenAI Functionalities to Bridge Novelty and Application}
% Our formative discussions highlighted the importance of framing AI as a collaborative assistant rather than an autonomous agent (\textbf{F2}--\textbf{F6}). While text-to-image GenAI is not positioned to replace radiologists (as it focuses on generating images rather than interpreting them), fears of AI replacing radiologists persist (\textbf{F4}, \textbf{F7}).
Our formative discussions highlighted the importance of understanding text-to-image GenAI's capabilities when coming up with potential applications for text-to-CT Scan GenAI. Radiologists were generally more familiar with traditional AI than text-to-image generative AI, which presented a barrier to imagining its applications. Initially, participants imagined use cases for text-to-text GenAI (\textbf{F2}) or traditional AI (\textit{i.e.}, image classification) (\textbf{F3}, \textbf{F4}, \textbf{F7}) before identifying scenarios relevant to text-to-image GenAI. As \textbf{F5} acknowledged: ``\emph{I’ve never been exposed to images generated by AI before},'' and struggled to envision practical applications, admitting ``\emph{honestly no ideas yet}.''

Providing radiologists with the opportunity to experience and experiment with text-to-image GenAI directly could help address this novelty and unfamiliarity before being asked about a very specific, complex application of text-to-image GenAI. Doing so can help bridge the gap between novelty and practical application and alleviate fears about AI’s role in radiology.

\begin{figure*}[t!]
    \centering
    \begin{subfigure}[t]{0.48\linewidth}
        \includegraphics[trim=1cm 5cm 2.3cm 4cm,clip,width=\linewidth]{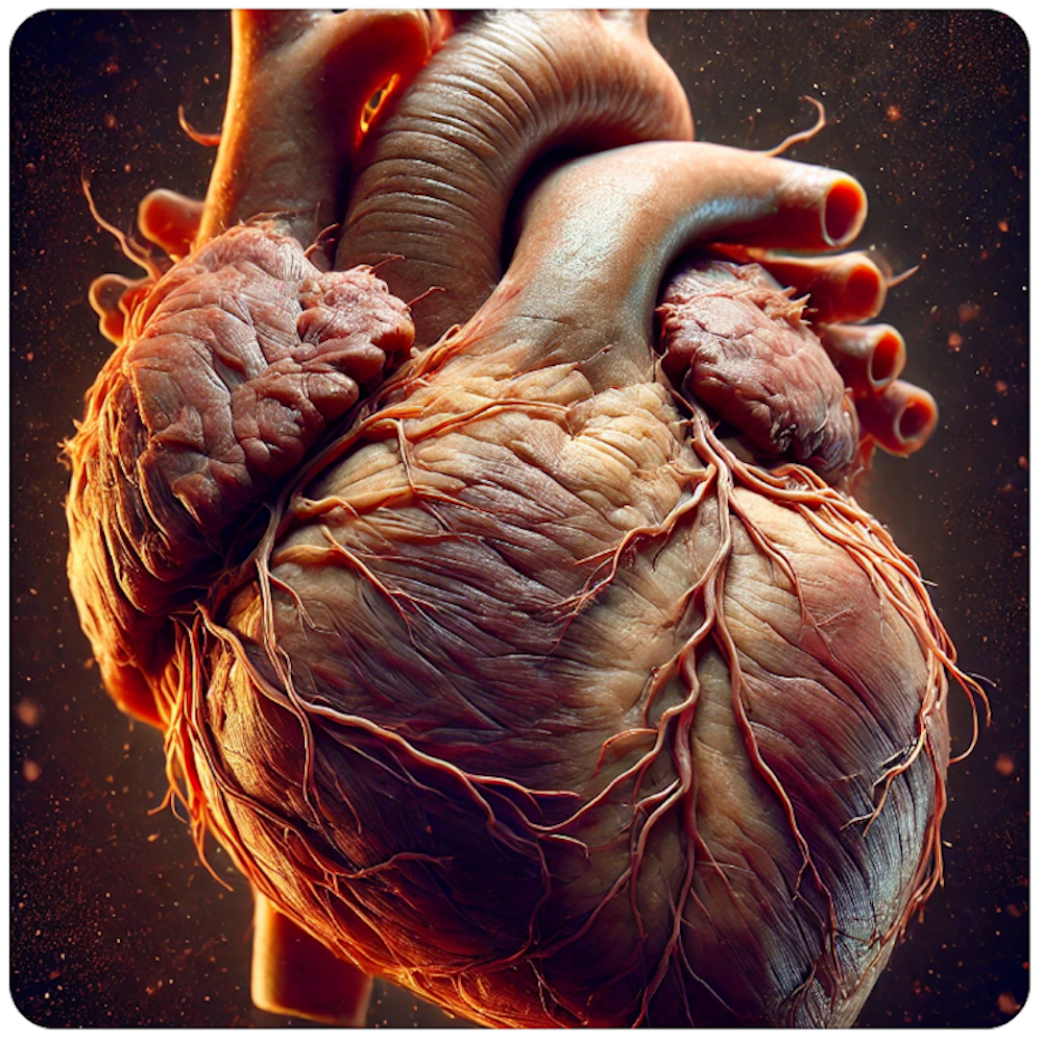}
        \caption{Generated image using ChatGPT. Prompt to generate was: Generate a realistic image of a heart, high-contrast, close-up.}
        \label{fig:heart-generated}
    \end{subfigure}
    \hfill
    \begin{subfigure}[t]{0.48\linewidth}
        \includegraphics[width=\linewidth]{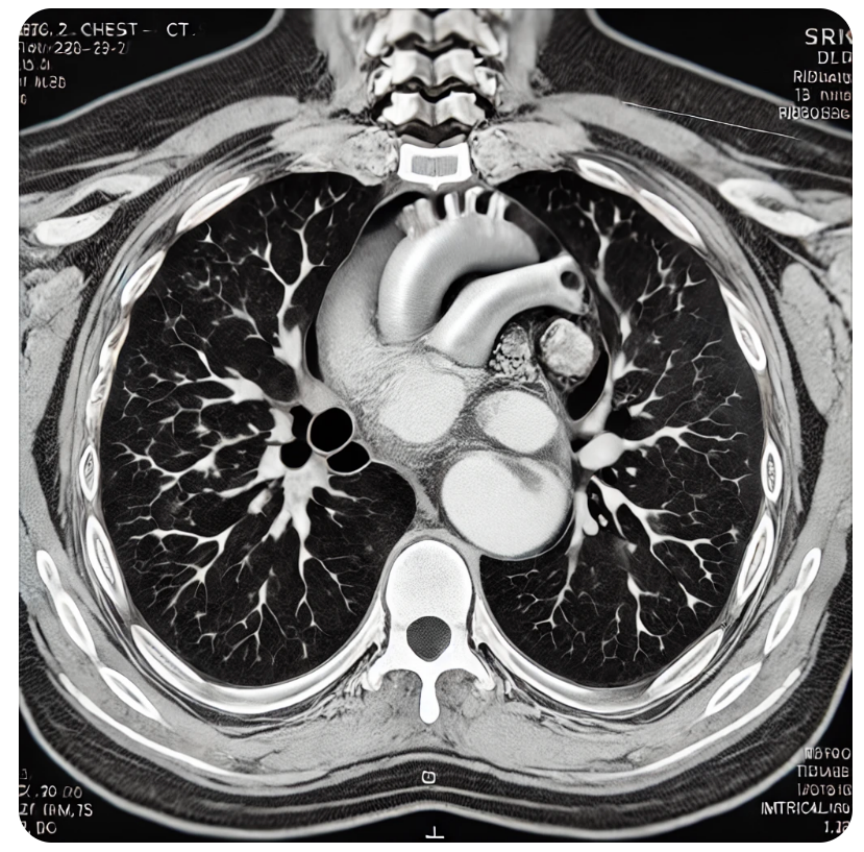}
        \caption{Generated image using ChatGPT. Prompt to generate was: Generate a realistic image of a chest CT scan.}
        \label{fig:chest-generated}
    \end{subfigure}
    \caption{Generated images from ChatGPT that participants were shown in the pre-probe discussion about text-to-image GenAI.}
    \label{fig:combined-figures}
\end{figure*}

\section{Semi-Structured Interviews}

\subsection{Pre-Probe Materials}
\label{pre-probe-materials}

Participants were shown generated medical images to help explain how text-to-image generation works. These images were only shown during the pre-probe portion of the interview to prepare the participant for seeing a more advanced text-to-image model in the design probe.

\subsection{Prompts \& Generations}

\begin{figure*}[p!]
    \centering
    \includegraphics[width=1\linewidth]{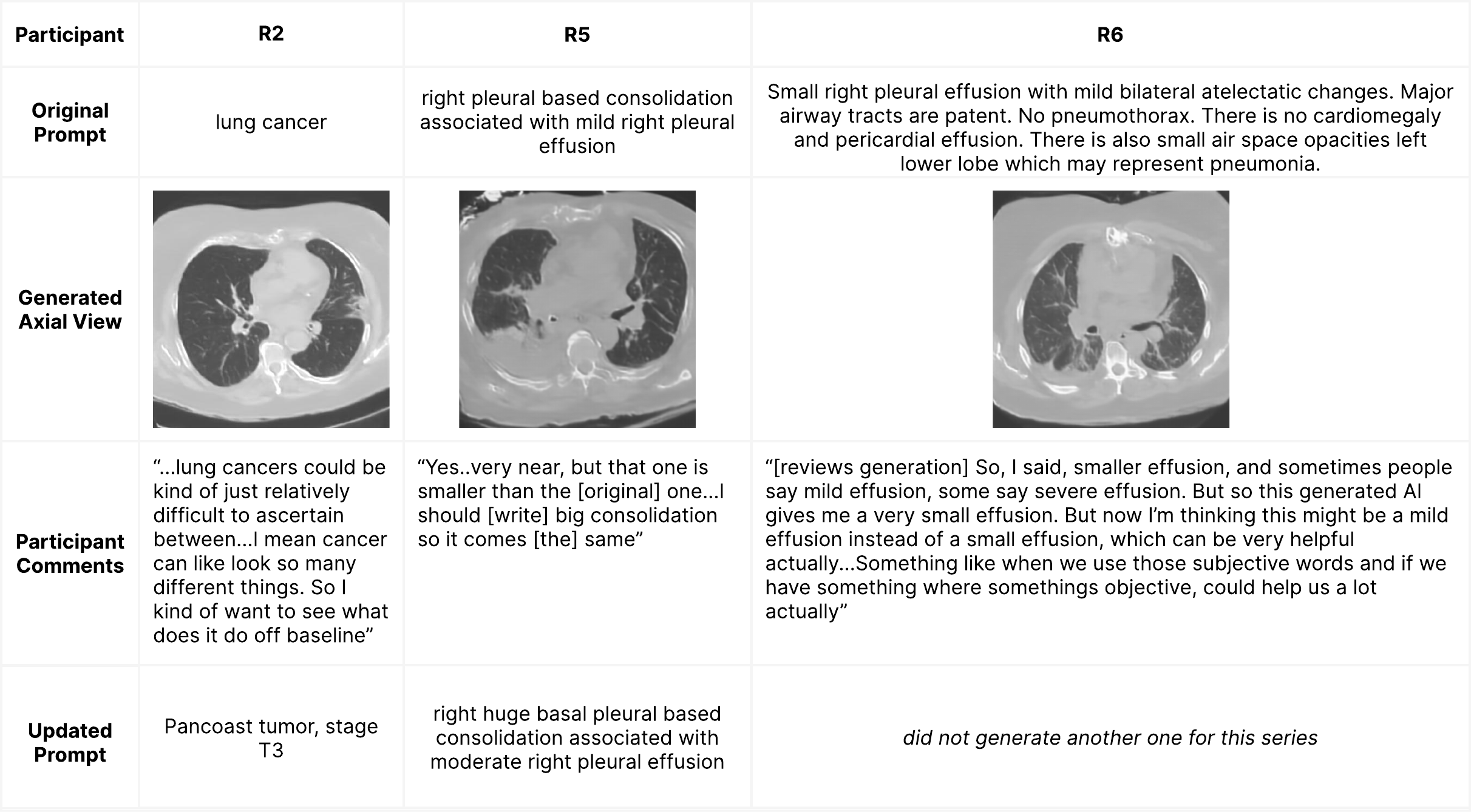}
    \caption{\textbf{Sample of Participants' Prompts \& Generated Outputs.} \textbf{R2} initially wanted to see how the model would generate general lung cancer and how it differentiates from a specific type of lung cancer. \textbf{R5} refined their prompt to be more specific in how the model generates the pathology. \textbf{R6} used their prompt as a way to validate their impression of the original patient CT, which they discovered needed to be adapted.}
    \label{fig:example-prompts}
\end{figure*}

\begin{figure*}[!p]
    \centering
    \includegraphics[width=1\linewidth]{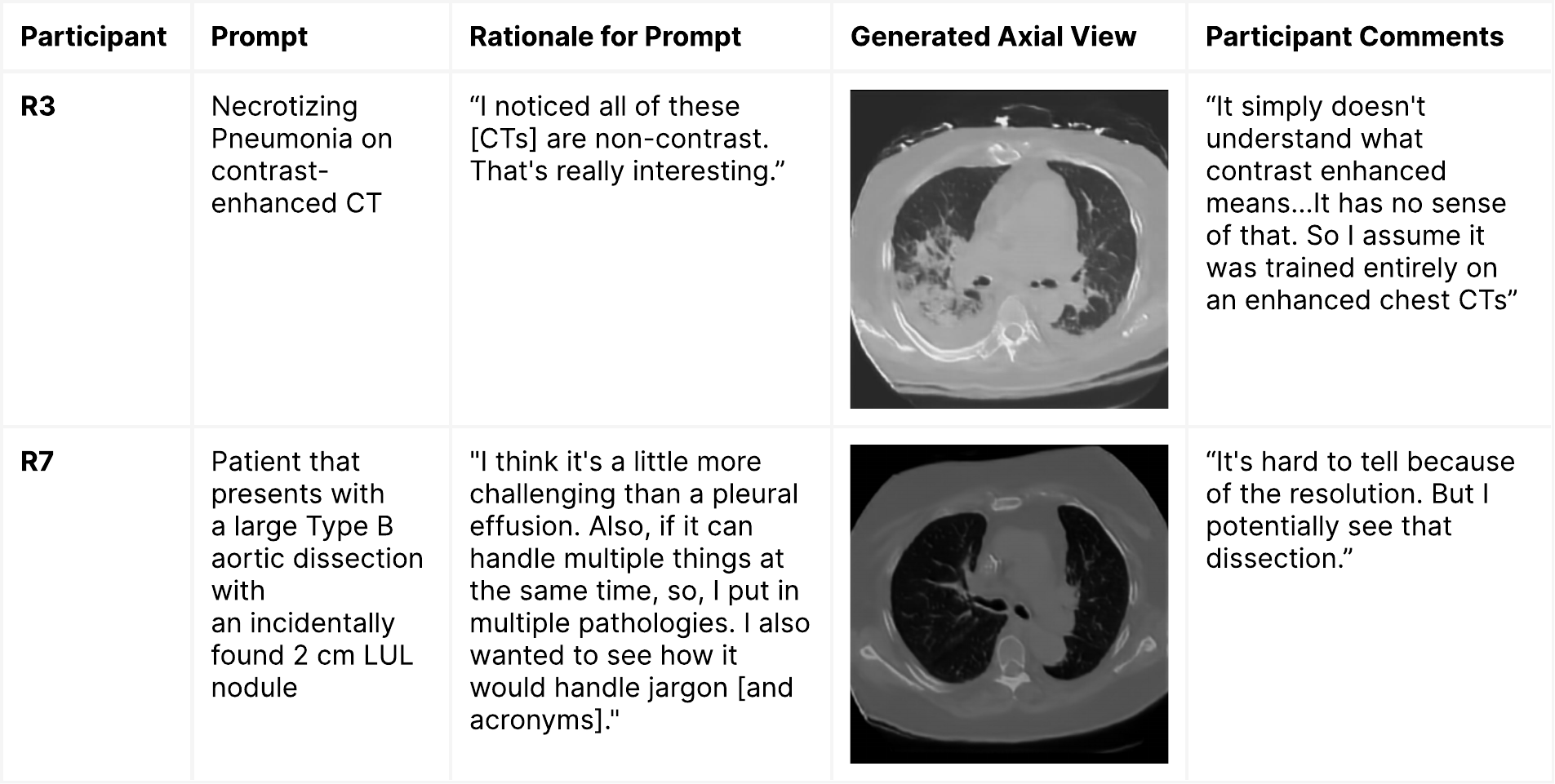}
    \caption{\textbf{Sample of Participants' Prompts \& Generated Outputs to Discover Technical Challenges.} \textbf{R3} chose this prompt to probe whether the model could actually generate a contrast-enhanced CT, which, as shown, cannot. \textbf{R7} chose this prompt to test multiple pathologies, more complicated pathologies, and jargon/acronyms. }
    \label{fig:example-prompts2}
\end{figure*}

An analysis of participants prompts reveal behaviors, such as auditing the models capabilities and limitations. In Figure~\ref{fig:example-prompts}, we show how participants updated their prompts after reviewing the generated image. Figure~\ref{fig:example-prompts2} reveals challenges of building medical T2I GenAI.

\begin{table*}[t]
    \centering
    \begin{tabularx}{\linewidth}{>{\centering\arraybackslash}m{2cm} >{\raggedright\arraybackslash}m{4cm} >{\raggedright\arraybackslash}X >{\arraybackslash}X}
        \toprule
        \textbf{Participant ID} & \textbf{Prompt} & \textbf{Comments about Prompt} & \textbf{Comments about Generation} \\
        \midrule
        R1 & necrotizing pneumonia & ``\textit{I was just curious to know how it's gonna show up on here. also, sometimes necrotizing pneumonia is. It's hard to differentiate with abscess. So I want to see, like how it differentiates like, or what it brings up for abscess versus necrotizing pneumonia.}'' & ``\textit{it's okay. It's good.}'' \\
        \midrule
        R2 & lung cancer & ``\textit{So I wanted to like. I wanted to start out vague to see kind of what it does with that before going to like something more specific, like like a mesothelioma that has like very specific features like it'll have like plaques on like the plural lining and stuff like that...I kind of want to see like what what does it do like off baseline.}'' & ``\textit{So this is our little kind of like, they're a little lesion right here. [pointing to a region on one of the slices]. cancer can absolutely look like this and like you know they made it kind of very like, more messy and obvious, which, like they gave it more aggressive features.
}'' \\
        \midrule
        R2 & Pancoast tumor; stage T3. & ``\textit{"pan coast lung tumor tumor is like, it's a very. It's going to be in a very specific region of the lung, and because it's located there, it's going to cause like specific symptoms to the patient and they're hard to see, because, like kind of where they're located, they can kind of hide sometimes behind the clavicles and stuff...we'll see what it does with that.}'' & ``\textit{Maybe this is what they wanted [pointing to a region in one of the slices]}''. \\
        \midrule
        R3 & bilateral IVC in otherwise normal chest CT & ``\textit{"It's an unusual diagnosis. I wouldn't expect in in a in a random data set. I don't know how many CTs this is based on. You know what it learned off of. If it is under 10,000, I wouldn't expect it to have encountered this. All of the words make sense. It will certainly be able to parse each of those words and make sense of them. But the combination is an extremely rare event.}'' & ``\textit{I should have asked it for contrast, but I could do that. They just had no idea what I was talking about.
It gave me an essentially normal chest CT}''. \\
        \midrule
        R3 & necrotizing pneumonia on contrast-enhanced CT & ``\textit{for a necrotizing pneumonia you really want a contrast.}'' & ``\textit{This is actually a better example of a necrotizing pneumonia than the one you had, though, is better. here I can definitely see the area of of neck. I see a vague area of pneumonia, and then a focal area of neck of cell death. So that is this is actually better}''. \\
        \bottomrule
    \end{tabularx}
    \caption{Participants' prompts, their comments about why they crafted that specific prompt, and their reactions to the generation for that prompt.}
    \label{tab:clue-violation}
\end{table*}

\clearpage

\begin{table*}[t]
    \centering
    \begin{tabularx}{\linewidth}{>{\centering\arraybackslash}m{2cm} >{\raggedright\arraybackslash}X >{\raggedright\arraybackslash}X >{\arraybackslash}X}
        \toprule
        \textbf{Participant ID} & \textbf{Prompt} & \textbf{Comments about Prompt} & \textbf{Comments about Generation} \\
        \midrule
        R4 & Moderate to severe right pleural effusion Partial right lung collapse Multiple liver masses One large liver cyst Ascites peritoneal thickening & [based on interpreting the provided real CT scan] & ``\textit{I can see the pleural effusion. Also the partial collapse...}'' \\
        \midrule
        R4 & Right lung hilar mass lots of smaller metastases bilaterally mild pleural effusion pleural masses reticular opacities and interlobular septal thickening in right lung right 7th rib erosion mediastinal lymphadenopaties & ``\textit{I tried to start with the most typical findings we see in a you know, stage 4 lung cancer to try to see how it works.}'' & ``\textit{Yeah we have mild pleural effusion...This one seems like a lymphedemopathy [pointing to a region in a slice]...Honestly, I expected a contrast enhanced CT not a non-contrast one. [the] generated [CT] has the [inlegible] as I expected in the right side pleural effusion.The mass might probably be here, but it is not. It is somehow obscured. There are some nodules, you know, I expected more prominent findings. I am not able to detect any bone lesions. It looks more like a Scapula fracture.}'' \\
        \midrule
        R4 & left rib fractures left bloody pleural effusion adjacent peripheral lung contusion small left pneumothorax & ``\textit{This is another typical case for, you know, a chest trauma.I tried to start with the most typical one, most classic finding.}'' & ``\textit{...working poorly in the bones part, But the mild pleural effusion, the contusion is what I had expected. So there's no pneumothorax.}'' \\
        \midrule
        R5 & right pleural based consolidation associated with mild right pleural effusion & [based on interpreting the provided real CT scan] & ``\textit{Very near, but that one is smaller...}'' \\
        \midrule
        R5 & right huge basal pleural based consolidation associated with moderate right pleural effusion & ``\textit{[I will] write big consolidation so it comes [up the] same}.'' & ``\textit{"See how it is very near now. I'm almost the same. Yes, it's almost the same like the original image you showed to me.}'' \\
        \midrule
        R5 & irregular speculated lesion in right apical region & ``\textit{Cause this one I know how it looks, so I want to see how it is...So if this one came as I used to see, this means this one [model] is very high percent [accurate].}'' & ``\textit{It's very small.}'' \\
        \midrule
        R5 & irregular speculated lesion 3 cm in right apicoposterior region & [size specification to lesion] & ``\textit{[Did] not generated this one}''. \\
        \midrule
        R6 & Small right pleural effusion with mild bilateral atelectatic changes. Major airway tracts are patent. No pneumothorax. There is no cardiomegaly and pericardial effusion. There is also small air space opacities left lower lobe which may represent pneumonia. & [based on interpreting the provided real CT scan] & ``\textit{So like I said, smaller effusion, and sometimes people say mild effusion, some say severe effusion. But so this generated AI gives me a very small effusion. But now I'm thinking this might be a mild effusion instead of a small effusion, which can be very helpful actually, because we also having similar problem with [incomprehensive], especially when you call it like large, small, mild, severe, significance. Something like when we use those subjective words and if we have something where something objective could help us a lot actually.}'' \\
        \bottomrule
    \end{tabularx}
    \caption{Participants' prompts, their comments about why they crafted that specific prompt, and their reactions to the generation for that prompt.}
    \label{tab:clue-violation}
\end{table*}

\clearpage

\begin{table*}[t]
    \centering
    \begin{tabularx}{\linewidth}{>{\centering\arraybackslash}m{2cm} >{\raggedright\arraybackslash}m{4cm} >{\raggedright\arraybackslash}X >{\arraybackslash}X}
        \toprule
        \textbf{Participant ID} & \textbf{Prompt} & \textbf{Comments about Prompt} & \textbf{Comments about Generation} \\
        \midrule
        R6 & Thickened bilateral basilar pleural may represent mesothelioma. There are also enlarged mediastinal lymph nodes measuring 2cm in short axis. There is also moderate right effusion. Mucus plug is noted on the left main bronchus associated with atelectatic changes & ``\textit{This is something like I don't see on the clinical setting usually. So I was just curious like what kind of images this AI is going to give me? Yeah, and like usually those things like we mentioned in our reports like as a negative findings like like we we usually say major. Airways are paid out and like, but here I said like there is a cost in the left main bronchus. So I just want to see what kind of images is is AI is going to give me because. As you can understand like. Being able to see those positive findings is more educational than commenting on negative findings.}'' & ``\textit{The quality seems little bit low. Is it because of my connection? Wow. Really cool. I have what I said.}'' \\
        \midrule
       R7 & Patient with large R pleural effusion of unknown etiology, with suspicious findings on the L side possibly consistent with atelectasis & ``\textit{I see all this fluid That's pretty obvious. I don't know what it's coming from, so I think, like the knowing the patient history and putting eyes on the patient might give a better idea, but potentially like something congested with the heart, potentially something congested with with the portal circulation. And then on the other side. This is all kind of weird to me. I don't know if that's fluid, or just like one consolidation, but that should, that shouldn't be there. I think I would use it comparatively to say, like, again, assuming that what it spits out is a realistic image, like, okay, like, this is what an adult. This should look like on the on the right side or on the left side. Rather. And compare that to what I'm seeing. Okay, well, if that's not what I'm seeing. What else could it possibly be?}'' & ``\textit{"this like very big right plural fusion that's pretty consistent with what we see on that the real image. And then over here it's hard to save. and then it gets into the question of like confirmation bias like, am I seeing what I'm actually seeing? Or am I seeing what I want to see, but I think to me like this is a pretty clear like consolidation over here.}'' \\
        
        \bottomrule
    \end{tabularx}
    \caption{Participants' prompts, their comments about why they crafted that specific prompt, and their reactions to the generation for that prompt.}
    \label{tab:clue-violation}
\end{table*}

\clearpage

\begin{table*}[t]
    \centering
    \begin{tabularx}{\linewidth}{>{\centering\arraybackslash}m{2cm} >{\raggedright\arraybackslash}m{4cm} >{\raggedright\arraybackslash}X >{\arraybackslash}X}
        \toprule
        \textbf{Participant ID} & \textbf{Prompt} & \textbf{Comments about Prompt} & \textbf{Comments about Generation} \\
\midrule
        R7 & Patient that presents with a large Type B aortic dissection with an incidentally found 2 cm LUL nodule & ``\textit{I have a pretty good idea of what an aortic dissection looks like on a Ct. So I'm curious to see how realistic that the report will be. I think it's a little more challenging than a plural fusion, because, like pleural fusion is just. you can tell that it's just fluid filling in. But aortic dissection. You have to find some way to Really demonstrate that there's a false lumen inside the aorta and then I also want to see if it'll. It'll also, if it can handle multiple things at the same time. So I put in multiple pathologies, aortic dissection. And then a lot of patients in real life will go in for something completely unrelated like they'll get a they'll go in for like stomach pain and find out they have to get their gallbladder out, and then also find out that they have this lung mass now on incidental imaging.}'' & ``\textit{So this is really cool. It's hard to tell because of the resolution. But I potentially see that dissection. I'm pointing on my screen. You can't tell what I'm looking at, but like the on the back wall. that dissection kind of going down. I don't know if that's what it's supposed to be. But that's potentially what I'm seeing.}'' \\
        \midrule
        R8 & Normal chest CT, with no presence of cardiomegaly, no presence of pericardial effusion, no presence of pleural effusion or consolidation& [NA] & ``\textit{"that's really cool. So this is not a real person? There's no particulate in the lung on the right side and the left side, the heart is much more normal. There's no consolidation and everything's clear on this one.}'' \\
        \midrule
        R8 & generate an type 1 abdominal aortic aneurysm & ``\textit{I just think it's very. It's like a very violent image, and I just want to I just want to see how ike what that's supposed to look like. And I think it's something they ask us to be able to observe pretty easily.}'' & ``\textit{[Compared to previous participant's generation for Type B aortic dissection] I also didn't ask for a dissection. Now that I think about it. I asked for an aneurysm, so that is probably a mistake on my end. Because it would probably be at the top of the aorta near the front. So I'm not sure how I would view that from this angle like it would probably need to be right around here. Maybe I'm not sure. I mean, I think that might have been a messaging error on my part, though.} \\
                \bottomrule
    \end{tabularx}
    \caption{Participants' prompts, their comments about why they crafted that specific prompt, and their reactions to the generation for that prompt.}
    \label{tab:clue-violation}
\end{table*}

\clearpage

\begin{table*}
    \centering
    \begin{tabularx}{\linewidth}{>{\centering\arraybackslash}m{2cm} >{\raggedright\arraybackslash}m{4cm} >{\raggedright\arraybackslash}X >{\arraybackslash}X}
        \toprule
        \textbf{Participant ID} & \textbf{Prompt} & \textbf{Comments about Prompt} & \textbf{Comments about Generation} \\
\midrule
        R9 & band-like atelectasis, prominent venous collaterals & ``\textit{I'm pretty sure there's Venus collaterals 'cause. I'm like, seeing pretty prominent veins up here by the skin. But maybe I'm wrong, so it'll help me by by letting me know I'm wrong. But I'm like seeing all these veins here. Feel like maybe this person has, like, SVC stenosis, a little bit And then with the band like Atelectasis, I was just, like kind of talking about this. So I guess I can compare it to what it shows me.}'' & ``\textit{So I guess my first thought is that I wasn't seeing enough contrast in the image. So feel like maybe this is the Venus collaterals it's trying to show me. I guess I'm also thinking that it's like a little bit smudgy. The heart anatomy is not clearly seen. It looks like they moved during. This is the band-Like atelectasis maybe? Or maybe I'm wrong...then there's like a little pleural effusion there}'' \\
        \midrule
        R9 & left lung pneumothorax & [generating an image finding not present in the provided real CT Scan] & ``\textit{No, it's not there.
Maybe it's like this tiny.
Maybe I'm not seeing it.
Maybe I'm a bad radiologist.}'' \\
                \bottomrule
    \end{tabularx}
    \caption{Participants' prompts, their comments about why they crafted that specific prompt, and their reactions to the generation for that prompt.}
    \label{tab:clue-violation}
\end{table*}

\end{document}